\documentclass[a4paper,12pt]{article} 

\usepackage{epsfig} 
\usepackage{amssymb} 
\usepackage[tbtags]{amsmath}
\usepackage{upref} 
\usepackage[normal]{caption}
\usepackage{float}
\usepackage{wrapfig}
\usepackage{subfigure}
\usepackage{longtable}
\def\asusy{$a_{\mu}^{{\rm \small SUSY}}$}
\def\bsg{$b\to s\gamma$}

\def\higgsu{m_{H_2}^2}
\def\higgsd{m_{H_1}^2}

\def\neumass{m_{\tilde\chi_1^0}}

\def\vev#1{\langle#1\rangle}

\def\higgsu{m_{H_u}^2}
\def\higgsd{m_{H_d}^2}
\def\higgsuew{m_{H_u}^2}
\def\higgsdew{m_{H_d}^2}

\def\neumass{m_{\tilde\chi_1^0}}
\def\charmass{m_{\tilde\chi_1^\pm}}
\newcommand{\crosssec}{\sigma_{\tilde\chi^0_1-p}}
\def\tanb{\tan\beta}

\def\neut{\tilde\chi_1^0}
\def\bsg{$b\to s\gamma$}

\def\asusy{a^{\rm SUSY}_\mu}

\def\relic{\Omega_{\tilde{\chi}_1^0}}

\newcommand{\captions}{\sf\caption}
\def\fig#1{Fig.\,\ref{#1}}
\def\eq#1{(\ref{#1})}

\def\lsim{\raise0.3ex\hbox{$\;<$\kern-0.75em\raise-1.1ex\hbox{$\sim\;$}}}
\def\gsim{\raise0.3ex\hbox{$\;>$\kern-0.75em\raise-1.1ex\hbox{$\sim\;$}}}

\begin{document}

\pagestyle{empty}

\rightline{DESY 04-034}
\rightline{FTUAM 04/06}
\rightline{IFT-UAM/CSIC-04-11}
\rightline{hep-ph/0405057}
\rightline{May 2004}   

\renewcommand{\thefootnote}{\alph{footnote}}

\vspace{1cm}
\begin{center}
  {\large{\bf
      Neutralino dark matter in supergravity theories with
      non-universal scalar and gaugino masses
      \\[5mm]
      }}
  \vspace{0.5cm}
  \mbox{
    \large{
      D.G.~Cerde\~no\,$^{1}$ and 
      C.~Mu\~noz\,$^{2}$
      }
    }
  \vspace{1cm}
  
  {\small
    {\it 
      ${^1}$ 
      II. Institut f\"ur Theoretische Physik, Universit\"at Hamburg,\\
      Luruper Chaussee 149, D-22761 Hamburg, Germany.\\
      \vspace*{2mm}
      \it $^2$ Departamento de F\'{\i}sica
      Te\'orica C-XI and Instituto de F\'{\i}sica
      Te\'orica C-XVI,\\ 
      Universidad Aut\'onoma de Madrid,
      Cantoblanco, 28049 Madrid, Spain.
      } 
    }
  
  \vspace{1cm}
  
  {\bf Abstract} 
  \\[7mm]
  
  \begin{minipage}[h]{14.0cm}
    We analyse the direct detection of neutralino dark matter in
    supergravity scenarios with non-universal soft scalar and gaugino
    masses.
    In particular, the 
    neutralino-nucleon cross section is computed and compared with the
    sensitivity of detectors. We take into account
    the most recent
    experimental and astrophysical constraints
    on the parameter space,
    including those coming from charge
    and colour breaking minima.
    Gaugino non-universalities
    provide a larger
    flexibility in the neutralino sector. In particular, when
    combined with non-universal scalars, 
    neutralinos close to the present 
    detection limits are possible with a wide range 
    of masses, from over $400$ GeV to almost $10$ GeV.
    We study the different possibilities which allow to increase or
    decrease the neutralino mass
    and explain the properties of those regions
    in the parameter space 
    with a large cross section.
  \end{minipage}
\end{center}

\newpage

\setcounter{page}{1}
\pagestyle{plain}
\renewcommand{\thefootnote}{\arabic{footnote}}
\setcounter{footnote}{0}

\section{Introduction}
\label{intro}

Weakly Interacting Massive Particles (WIMPs) are
plausible candidates for the dark matter in the Universe \cite{mireview}.
They are specially interesting because they can be present in the
right amount to explain the matter density observed in
the analysis of galactic rotation curves \cite{Persic}, cluster of
galaxies
and large scale flows \cite{Freedman}, 
$0.1\lsim \Omega\, h^2\lsim 0.3$ 
($0.094\lsim\Omega\, h^2\lsim 0.129$
if we take into account the recent data obtained by the
WMAP satellite \cite{wmap03-1}).

The leading candidate for WIMP is 
the lightest neutralino \cite{mireview}, $\neut$,
a particle predicted by the
supersymmetric (SUSY) extension of the standard model. These neutralinos
are usually stable and therefore may be left over from the
Big Bang.
Thus they 
will cluster gravitationally with ordinary stars in the galactic halos, 
and 
in particular they will be present in our own galaxy, the Milky Way.
As a consequence there will be a flux of these dark matter particles
on the Earth.

Many underground 
experiments have been carried out around the world in order to detect this
flux, by observing the elastic scattering of the
dark matter particles on target nuclei through nuclear 
recoils \cite{mireview}. 
In fact, one of the current experiments, the DAMA collaboration, has
reported data favouring the existence of a WIMP signal \cite{dama}.
Taking into account uncertainties on the halo model,
it was claimed that 
the preferred range of the WIMP-nucleon cross section
is $\sigma \approx 10^{-6}-10^{-5}$ pb 
for a WIMP mass smaller than $500-900$ GeV \cite{dama,halo}.
Unlike this spectacular result, other collaborations such as 
CDMS \cite{experimento2} and EDELWEISS \cite{edelweiss}, 
claim to have excluded important regions of the DAMA 
parameter space.

In any case, due to these and other projected experiments
\cite{mireview}, it seems very plausible that the dark matter 
will be found in the near future. For example,
GEDEON \cite{IGEX3}  
will be able to explore
positively a WIMP-nucleon 
cross section
$\sigma \gsim 3\times 10^{-8}$ pb.
Similarly, CDMS Soudan (an expansion of the CDMS experiment
in the Soudan mine), 
will be able to test 
$\sigma \gsim 2\times 10^{-8}$ pb.
But the most sensitive 
detector will be 
GENIUS \cite{HDMS2}, 
which will be able to test a WIMP-nucleon cross section
as low as $\sigma\approx 10^{-9}$ pb.

Given this situation, and assuming that the dark matter 
is a neutralino, it is natural to wonder how big 
the cross section for its direct detection can be.
Obviously,
this analysis is crucial in order to know the
possibility of detecting dark matter 
in the experiments.
In fact, the analysis of the neutralino-proton cross section 
has been carried out by many authors and during many 
years \cite{mireview}.
The most recent studies take into account the present
experimental and astrophysical constraints
on the parameter space. 
Concerning the former, 
the lower bounds on the Higgs mass
and the supersymmetric particles,
the $b\to s\gamma$ branching ratio, and the
supersymmetric contribution to the muon anomalous magnetic moment,
$\asusy$, 
have been considered.
The astrophysical bounds on the matter density
mentioned above have also been
imposed on the 
theoretical computation of the relic neutralino density,
assuming thermal production.
In addition, 
the constraints that the absence of dangerous charge
and colour breaking minima imposes on the parameter space
have also been taken into account \cite{cggm03-1}.

In the usual minimal supergravity (mSUGRA) scenario, where the soft terms
of the minimal supersymmetric standard model (MSSM) 
are assumed to be universal at the unification scale, 
$M_{GUT} \approx 2\times 10^{16}$ GeV,
and radiative electroweak symmetry
breaking is imposed, 
the cross section turns out
to be constrained by
$\sigma_{\tilde{\chi}_1^0-p}\lsim 3\times 10^{-8}$ pb \cite{mireview}.
Clearly, in this case, present experiments are not sufficient and
more sensitive detectors
producing further data 
are needed.

The above result can be modified by taking into account
possible departures from the mSUGRA scenario, and different
possibilities
have been proposed in the literature. For example, 
when the GUT condition is relaxed and 
an intermediate scale is allowed,
the cross section increases significantly \cite{intermediate}.
However,
the experimental bounds
impose
$\sigma_{\tilde{\chi}_1^0-p}\lsim 4\times 10^{-7}$ pb.
And, in fact, at the end of the day, the preferred astrophysical range
for the relic neutralino density, 
$0.1\leq\relic\,h^2\leq 0.3$,
imposes 
$\sigma_{\tilde{\chi}_1^0-p}\lsim  10^{-7}$ pb, i.e.,
beyond the sensitivity of present experiments
\cite{cggm03-1}.

A more general situation in the context
of SUGRA than universality, the presence of non-universal
soft scalar 
\cite{Fornengonew,Arnanew,Bottino,arna2,Arnowitt,Santoso,Drees,Nojiri,darkcairo,Arnowitt3,nosopro,Dutta,Rosz,Profumo,cggm03-1,Farrill}
and 
gaugino 
masses \cite{Nath2,darkcairo,nosopro,Orloff,Dutta,Birkedal,Roy2,cggm03-1}
has also been considered. 
Non-universalities in both the scalar and gaugino sectors
were also studied in \cite{pallis} in the context of a SUSY GUT inspired
MSSM version.
In particular, for some special choices of the non-universality
in the scalar sector the cross section can be increased
significantly with respect to the universal scenario,
and 
allowed by all experimental and
astrophysical 
constraints. In fact,
not
only large regions of the parameter space are accessible for
future experiments, but also in part of them
the sensitivity of present experiments is 
reached,
$\sigma_{\tilde{\chi}_1^0-p}\approx 10^{-6}$ pb (for a recent analysis, 
see e.g. Ref.~\cite{cggm03-1})\footnote{This
  is similar to what occurs in the so-called
  effMSSM scenario 
  \cite{bg95,Bottinoeff,bednyakov,mandic,baltzg02,knrr02,eoss03-1},
  where the parameters are defined directly at the electroweak scale.
  }.
On the other hand, non-universality in the gaugino sector 
also increases the cross section. However, the above sensitivity region
cannot be reached, and 
$\sigma_{\tilde{\chi}_1^0-p}\lsim  10^{-7}$ pb.

The aim of this paper
is to investigate the general case, where non-universalities are
present both in the scalar and gaugino sectors, and
to carry out a detailed analysis of 
the prospects
for the direct detection of neutralino dark matter in these
scenarios.
In this analysis we will take into account the present
experimental and astrophysical constraints mentioned above,
as well as the constraints coming from charge and colour breaking minima.
In the light of the recent experimental results,
we will be specially interested in studying how big the cross section
can be.
Our purpose is to provide a general analysis which can be used in the
study of any concrete model.

The paper is organised as follows.
In Section~\ref{departures} we will discuss the situation concerning the
neutralino-proton cross section in SUGRA theories. In particular, 
we will review the possible departures from mSUGRA, with either
non-universal soft scalar masses or soft gaugino masses, which give
rise to large values of the cross section.
In Section~\ref{general} we will study the general case where both
scalar and gaugino non-universalities are present. We will indicate
the conditions under which a significant enhancement of the resulting
cross section is obtained. Finally, the conclusions are left for
Section~\ref{conclusions}.

\section{Departures from the mSUGRA scenario}
\label{departures}

In this section we will review possible departures from
the mSUGRA scenario and their impact on the neutralino-proton
cross section. Let us first recall that in mSUGRA
one has only four free parameters defined at the GUT scale:
the soft scalar mass $m$, the soft gaugino mass $M$, 
the soft trilinear coupling $A$, and the ratio of the Higgs vacuum
expectation values, $\tan\beta\equiv\vev{H_u^0}/\vev{H_d^0}$.
In addition the sign of the Higgsino mass parameter, $\mu$,
remains also
undetermined by the minimization of the Higgs potential, which implies
\begin{equation}
  \mu^2 = \frac{\higgsdew - \higgsuew \tan^2 \beta}{\tan^2 \beta -1 } - 
  \frac{1}{2} M_Z^2\,.
  \label{electroweak}
\end{equation} 
Using these parameters the neutralino-proton cross section
has been analysed exhaustively in the literature, as mentioned
in the Introduction.
Taking into account all kind of experimental and astrophysical
constraints, the result is that the scalar cross section is bounded to be
$\sigma_{\tilde{\chi}_1^0-p}\lsim 3\times 10^{-8}$ pb 
(for a recent analysis, 
see e.g. Ref.~\cite{cggm03-1}).
Obviously, in mSUGRA, present experiments for the 
direct detection of dark matter are not sufficient and
more sensitive detectors
producing further data 
are needed.

The neutralino-proton cross section can be increased in different
ways when the structure of mSUGRA for the soft terms is abandoned. 
In particular, it is possible to enhance the scattering
channels involving exchange of CP-even neutral Higgses by reducing the
Higgs masses, and also by increasing the Higgsino components of
the lightest neutralino.
A brief analysis based on the Higgs
mass parameters, $\higgsdew$ and $\higgsuew$, at the electroweak scale
can clearly show how these effects can be achieved.

First, a decrease in the values of the Higgs masses can be
obtained by increasing $\higgsuew$ (i.e., making it less negative) 
and/or decreasing $\higgsdew$. 
More specifically, the value of the mass of the heaviest CP-even
Higgs, $H$, can be very efficiently lowered under these
circumstances.
This is easily understood by analysing 
the (tree-level) 
mass of the CP-odd Higgs\footnote{The CP-odd Higgs mass generically
  receives very small  
  one-loop corrections, of order 1\%. 
  For this reason we will only consider its tree-level value in the 
  discussion.}, 
$A$,
\begin{equation}
  m^2_A=m_{H_d}^2+m_{H_u}^2+2\mu^2\,,
  \nonumber
  \label{ma} 
\end{equation}
which can be rewritten as
\begin{equation}
  m^2_A\approx m_{H_d}^2-m_{H_u}^2-M_Z^2\,,
  \label{ma2} 
\end{equation}
taking into account that,
for reasonably large values of
$\tan\beta$, expression \eq{electroweak} can be approximated as
\begin{equation}
  \mu^2\approx -\higgsuew-\frac{1}{2} M_Z^2\,.
  \label{electroweak2}
\end{equation}
Since the heaviest CP-even Higgs,
$H$, is almost degenerate in mass with $A$,
lowering $m^2_A$ we obtain a decrease in $m_H$
which produces an increase in the scattering channels through Higgs
exchange\footnote{
  Let us remark that this is true for values of $m^2_A$ above a
  certain critical mass (which corresponds to the
  intense-coupling regime for the Higgses \cite{intense} 
  and also sets the 
  maximum value of the lightest Higgs mass). 
  For values of $m^2_A$ below this critical mass, $m_H$ is stabilised 
  close to its minimal value and it is now 
  the mass of the lightest Higgs, $h$,
  which decreases with decreasing $m^2_A$, 
  thus obtaining a further increase in the cross section.
  This can occur, e.g., in the case of very light neutralinos, as we
  will see in Section \ref{sec_verylight}.
  }.

Second, through the increase in the value of
$\higgsuew$ an increase in the Higgsino components of
the lightest neutralino can also be achieved. 
Making $\higgsuew$ less negative, its positive contribution to 
$\mu^2$ in \eq{electroweak2} would be smaller.
Eventually $|\mu|$ will be of the order of
$M_{1}$, $M_{2}$ and 
$\tilde{\chi}_1^0$ will then be a mixed Higgsino-gaugino state.
Thus scattering channels through Higgs exchange become more important
than in mSUGRA, where $|\mu|$ is large and $\tilde{\chi}_1^0$ is mainly
bino.
It is worth emphasizing however that
the effect of lowering the Higgs masses is typically more important,
since it can provide large values for the neutralino-nucleon cross
section even in the case of bino-like neutralinos.

\subsection{Non-universal scalars}
\label{sec_nunivsc}

Non-universal soft parameters can produce the above mentioned
effects. Let us first consider non-universalities in the scalar
masses \cite{Fornengonew,Arnanew}.
We can parameterise
these in the Higgs sector, at the GUT scale, as
follows:
\begin{equation}
  m_{H_{d}}^2=m^{2}(1+\delta_{1})\ , \quad m_{H_{u}}^{2}=m^{2}
  (1+ \delta_{2})\ .
  \label{Higgsespara}
\end{equation}
Concerning squarks and sleptons we will assume
that the three generations have the
same mass structure:
\begin{eqnarray}
  m_{Q_{L}}^2&=&m^{2}(1+\delta_{3})\ , \quad m_{u_{R}}^{2}=m^{2}
  (1+\delta_{4})\ , 
  \nonumber\\
  m_{e_{R}}^2&=&m^{2}(1+\delta_{5})\ ,  \quad m_{d_{R}}^{2}=m^{2}
  (1+\delta_{6})\ , 
  \nonumber\\
  m_{L_{L}}^2&=&m^{2}(1+\delta_{7})\ .    
  \label{Higgsespara2}
\end{eqnarray}
Such a structure avoids potential problems
with flavour changing neutral 
currents\footnote{Another possibility would be to assume 
  that the first
  two generations have the common scalar mass $m$, and
  that non-universalities are allowed only for the third generation.
  This would not modify our analysis since, as we will see below, 
  only the third generation is relevant in our
  discussion.}.
Note also that whereas $\delta_{i} \geq -1 $, $i=3,...,7$, in order to
avoid an unbounded from below (UFB) direction breaking charge and
colour, 
$\delta_{1,2} \leq -1$
is possible as long as 
$m_1^2\,=\,m_{H_{d}}^2+\mu^2>0$ and $m_2^2\,=\,m_{H_{u}}^2+\mu^2>0$  
are fulfilled.

An increase in $\higgsu$ at the electroweak scale can be obviously
achieved by 
increasing its value at the GUT scale, i.e., with the choice
$\delta_2>0$. 
In addition, this is also produced when $m_{Q_{L}}^2$ and $m_{u_{R}}^2$
at $M_{GUT}$ decrease,
i.e. taking $\delta_{3,4} < 0$, due to 
their (negative) contribution proportional
to the top Yukawa coupling in the renormalization group equation (RGE)
of $m_{H_u}^2$.

Similarly, a decrease in the value of $\higgsd$ at the electroweak
scale can be 
obtained by decreasing it at the GUT scale with $\delta_1<0$. Also,
this effect is produced when $m_{Q_{L}}^2$ and $m_{d_{R}}^2$ 
at $M_{GUT}$ increase, due to their (negative) contribution proportional
to the bottom Yukawa coupling in the RGE
of $m_{H_d}^2$.
Thus one can deduce that $m^2_A$ 
will be reduced 
by choosing also $\delta_{3,6} > 0$.

In fact non-universality in the Higgs sector gives the most important
effect, and including the one in the sfermion sector the cross
section only increases slightly. Thus in what follows we will take
$\delta_{i}=0$, $i=3,...,7$.

Taking into account this analysis, several scenarios were discussed
in Ref.~\cite{cggm03-1}, obtaining that large values for the 
cross section are possible. For example, with
$\delta_{1}=0,\, \delta_2=1;\ 
\delta_{1}=-1,\, \delta_2=0;\
\delta_{1}=-1,\, \delta_2=1$,
one obtains regions of the parameter space accessible for
experiments\footnote{
  Note in this sense that varying the soft Higgs masses, $\higgsd$ and
  $\higgsu$, corresponds to varying $\mu$ and $m_A$ arbitrarily in the
  effMSSM scenario.
  }. 
Interestingly, it was also realised that these choices were helpful in
order to prevent the appearance of UFB minima in the Higgs potential.

The neutralino mass in these cases has a lower limit which can be
derived from the effect of the experimental constraints 
on the common gaugino mass, $M$, and the $\mu$
parameter. Small values of $M$ are restricted by the 
constraints on the Higgs mass and $\asusy$, and by 
\bsg. The latter becomes very important for large values of 
$\tan\beta$. These imply $M\gsim200$ GeV at
the GUT scale and thus $M_1\gsim80$ at the electroweak scale, which
can be interpreted as a lower bound for the mass of a 
bino-like neutralino. 
Similarly, the value of the $\mu$ parameter is restricted by 
the lower bound on the lightest chargino, thus having $|\mu|\gsim 100$
GeV.
Although this would set a lower constraint on Higgsino-like
neutralinos, these give rise to very small
relic densities and are therefore further restricted.
For these reasons, the neutralino mass in SUGRA theories with only
non-universal scalars cannot be arbitrarily lowered.

\subsection{Non-universal gauginos}
\label{sec_nunivg}

Let us now review the effect of the non-universality in the gaugino
masses. We can parameterise this as follows:
\begin{eqnarray}
  M_1=M\ , \quad M_2=M(1+ \delta'_{2})\ ,
  \quad M_3=M(1+ \delta'_{3})
  \ ,
  \label{gauginospara}
\end{eqnarray}
where $M_{1,2,3}$ are the bino, wino and gluino masses, respectively,
and $\delta'_i=0$ corresponds to the universal case.

In order to increase the cross section
it is worth noticing that
$M_3$ appears in the RGEs of squark masses.
Thus the contribution of squark masses
proportional
to the top Yukawa coupling in the RGE of $m_{H_u}^2$ will
do this less negative if $M_3$ is 
small.
As discussed above, this produces an enhancement in the cross section.

Because the mass of the lightest Higgs is very dependent on the value
of $M_3$, its decrease is very limited. 
In fact, in order
to satisfy the lower limit of $M_3$, $M$ in \eq{gauginospara} may have
to increase, thus rather than a decrease in $M_3$ what
one obtains is an effective increase in $M_1$ and $M_2$,
which
leads to a larger (less
negative) value of $\higgsu$ and thus a reduction in the value of
$|\mu|$.
This in turn
implies heavier neutralinos, 
when the lightest neutralino is mostly
gaugino, and an increase of the Higgsino composition, which would be
dominant if $M_1>|\mu|$ at the electroweak scale.
For this reason there is a slight raise in the predictions for
$\crosssec$\footnote{Note that the value of $\higgsd$ also increases,
  thus $m_A$ calculated from \eq{ma2} is typically
  not very affected.}.
Finally, decreasing the ratio $M_3/M_1$ leads to a more efficient
neutralino annihilation due to the enhancement in the Higgsino
components of $\neut$, entailing a reduction of $\relic$.
An example with $\delta'_{2}=0, \delta_3=-0.5$, producing an
increase in the dark matter cross section with respect to the
universal case, can be found in Ref.~\cite{cggm03-1}, where it was
also argued that this choice of gaugino non-universalities is
good to avoid UFB constraints.

On the other hand, 
increasing the value of $M_3$ with $\delta_3>0$ presents the advantage
that the constraint on the lightest Higgs mass is more easily
fulfilled. 
Equivalently, this implies that the value of $M$ in \eq{gauginospara}
can be lowered and thus have an effective decrease in $M_1$ (and
also $M_2$ unless $\delta_2>0$ is chosen).
This makes it possible to obtain lighter neutralinos with a larger
bino composition,
satisfying all the experimental and astrophysical constraints. 
However, because of the above arguments the values of the
cross section would slightly decrease with respect to the universal
case.
Despite the decrease in the neutralino mass, the appearance of light
neutralinos in this case is also restricted by 
the results on the relic density. In particular, very light
neutralinos typically give rise to a very
large $\relic$, which would be 
incompatible with present observations. A reduction in the
relic density would only be obtained along the narrow
resonances with the
lightest Higgs and the $Z$ at $\neumass=m_h/2,\,M_Z/2$, respectively,
thus setting the lower bound for the neutralino mass in these 
scenarios with only gaugino non-universalities.
We will later come back to this point in the context of a more general
SUGRA scenario.

The main role of $M_2$ is altering the lightest neutralino
composition. 
It is well known that decreasing the ratio $M_2/M_1$, thus increasing
the wino component of the lightest neutralino, enhances the neutralino
detection rates and provides a more effective neutralino annihilation
through channels mediated by $\tilde\chi_2^0$ and $\tilde\chi_1^+$ and
coannihilations with these
\cite{mizuta,birkedal01,Orloff}.
However, 
this is only effective 
when $M_2/M_1\lsim 0.5$ (which leads to $M_2\lsim M_1$ after
the running from the GUT scale in the MSSM),
and as pointed out in Refs.~\cite{mizuta,Orloff}, as soon as the wino
component begins to dominate, the resulting relic
density becomes too small. 
Variations in the value of $M_2$ also affect the
predictions for $\asusy$. For instance,
decreasing $M_2$, the contribution of the diagrams involving
intermediate chargino-sneutrino states to $\asusy$
becomes more important and it may increase beyond its upper
bound. This sets a more stringent lower bound on the masses of the
neutralino. If, on the other hand, $M_2$ is increased, the decrease in
$\asusy$ will set a stronger upper constraint on $\neumass$.

Summarising, although gaugino non-universalities also alter the
predictions for the neutralino-nucleon cross section, their influence
for raising it is not as 
important as the one arising from non-universal scalars.
In particular, none of the above choices for the parameters allows
the appearance of neutralinos in the detection range of present dark
matter experiments.

\section{General case: non-universal scalars and gauginos}
\label{general}

In this Section we will consider the general case where the soft
supersymmetry-breaking terms for both scalar and
gauginos have a non-universal structure.
Analysing the effect of combining these non-universalities is
interesting from the theoretical point of view, since such a
structure can be recovered in the low-energy limit of some
phenomenologically appealing string
scenarios.
For example, D-brane constructions in Type I string 
possess this property \cite{Rigolin} when the gauge group of the
Standard Model originates from different stacks of D-branes.

We will be mostly interested in analysing the conditions under which
high values for the cross section are obtained. For this reason, we
will concentrate on some interesting choices for scalar
non-universalities, 
exemplified by the following cases \cite{cggm03-1}
\begin{eqnarray}
  &&a)\quad\delta_{1}=0,\quad\ \  \delta_2=1;\nonumber\\ 
  &&b)\quad\delta_{1}=-1,\quad \delta_2=0;\nonumber\\ 
  &&c)\quad\delta_{1}=-1,\quad \delta_2=1,
  \label{nunivsc}
\end{eqnarray}
and study the effect of adding gaugino non-universalities to these.

The soft terms are given at a high energy scale which in our
analysis will be taken to be the GUT scale,
where unification of the 
gauge coupling constants takes place. In our
computation the most recent experimental and astrophysical
constraints will be taken into account. In particular, the lower
bounds on the masses of the supersymmetric particles and on the
lightest Higgs have been implemented, as well as the experimental
bounds on the branching ratio of the \bsg\ process and on 
$\asusy$.  The evaluation of the neutralino relic density is carried
out with the program {\tt micrOMEGAs} \cite{micro}, and, due to its
relevance, the effect of the WMAP constraint on it will be shown
explicitly. Finally,
dangerous charge and colour breaking minima of the Higgs potential
will be avoided by excluding UFB directions.

Concerning $\asusy$, we have taken into account the
recent experimental result for the muon
anomalous magnetic moment \cite{g-2}, as well as the most recent
theoretical evaluations of the Standard Model contributions
\cite{newg2}. It is found that when $e^+e^-$ data
are used the experimental excess in $(g_\mu-2)$ would constrain a
possible supersymmetric contribution to be
$\asusy=(27.1\,\pm\,10)\times10^{-10}$. In our analysis we will impose
consistency with this value at $2\sigma$ level and thus use the
constraint\footnote{
 It is worth
 noticing at this point that when tau data are used a smaller
 discrepancy with the experimental measurement is found.}
$7.1\times10^{-10}\lsim\asusy\lsim47.1\times10^{-10}$.
For details on how the rest of the experimental bounds are implemented 
see \cite{cggm03-1}.

The parameter space consists of a common scalar mass,
$m$, with the non-universal Higgs masses given by \eq{Higgsespara}
and the three choices 
\eq{nunivsc}, a
common trilinear parameter, $A$, and a gaugino sector which can be
specified with the three independent parameters, $M$, $\delta'_2$ and
$\delta'_3$, in
\eq{gauginospara}.
The set of inputs is completed with $\tan\beta$ and
the sign of the $\mu$ parameter.

Because the sign of $\asusy$ is basically given by $\mu\,M_2$,
we will consider $sign(M_2)=sign(\mu)$ in order to fulfil the
experimental result\footnote{
 Note that if the constraint on $\asusy$ resulting from tau data is
 taken into account, 
 a different sign for $M_2$ and $\mu$ could in principle also be
 used. 
 Nevertheless, in order to reproduce the negative values of $\asusy$,
 which are very small in modulus, 
 very large values of $|M_2|$ are necessary.
 This possibility is therefore very constrained.
 }. 
Similarly, the constraint on the \bsg\ branching
ratio is much weaker when $sign(M_3)=sign(\mu)$. 
Finally, variations in the sign of $M_1$ do not induce significant
changes in the allowed regions of the parameter space (e.g., its
effect on $\asusy$, due to diagrams with neutralino intermediate states,
is smaller than the one of $M_2$). However, when
$sign(M_1)=sign(\mu)$ the theoretical predictions for $\crosssec$ are
larger. 
For these reasons we will restrict our analysis to positive values of
$M_{1,2,3}$ and $\mu>0$. 
Note in this sense, that due to the
symmetry of the RGEs, the results for ($M_{1,2,3},\mu,A$) are
identical to those for ($-M_{1,2,3},-\mu,-A$).

Due to the importance of the gluino mass parameter, 
we will group the possible gaugino 
non-universalities in two different cases,
depending on whether the
ratio $M_3/M_1$ at the GUT scale 
decreases or increases with respect to its value in 
the universal
case, and analyse variations of $M_2$
within each case.

\subsection{Decrease in $M_3/M_1$}
\label{decm3m1}

Let us first study the consequences of decreasing the value
of $M_3$ with respect to $M_1$ as a complement 
to the scalar non-universalities \eq{nunivsc}. We will therefore
choose $\delta'_3<0$ in \eq{gauginospara}.
In order to satisfy the constraint
on the lightest Higgs mass,
higher values of $M$, and therefore of
$M_1$ are
necessary. In those cases where the lightest 
neutralino is mostly bino this implies that the neutralino mass is
increased. Thus it is possible to find heavier neutralinos with a
relatively high value for their direct detection cross section.

\begin{figure}
  \hspace*{-1.5cm}\epsfig{file=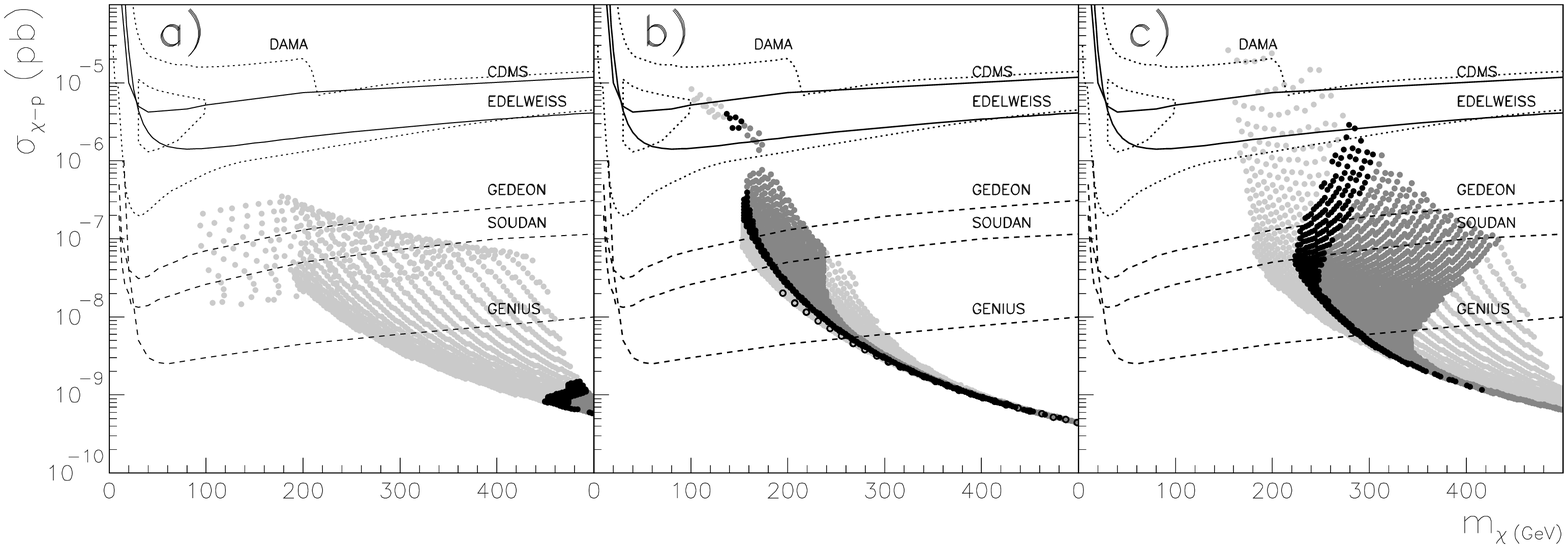,width=18cm}
  \captions{
    Scatter plot of the 
    scalar neutralino-proton cross section $\crosssec$ as a function
    of the neutralino mass $\neumass$ for
    $\delta'_{2,3}=-0.25$ and the three choices for
    non-universal scalars \eq{nunivsc} in a case with $\tan\beta=35$
    and $A=0$. The light grey dots correspond to points fulfilling all
    the experimental constraints. The dark grey dots represent points
    fulfilling in addition 
    $0.1\leq\relic\,h^2\leq0.3$ and the black ones correspond to those
    consistent with
    the WMAP range. 
    Points excluded by the UFB constraints are represented with
    circles.
    The sensitivities of present and projected experiments are
    also depicted with solid and dashed lines, respectively.
    The large (small) area bounded by dotted lines is allowed by the
    DAMA experiment when astrophysical uncertainties are (are not)
    taken into account.
    }
  \label{cross35n}
\end{figure}

Regarding $M_2$, 
let us begin by considering also a reduction in $M_2/M_1$, 
by taking 
$\delta'_2<0$ in \eq{gauginospara}. The gaugino structure at the GUT
scale would therefore be $M_1>M_2\sim M_3$.
An example with $\delta'_{2,3}=-0.25$ is shown
in \fig{cross35n}, where the neutralino-nucleon cross section is
plotted 
versus the neutralino mass, $\neumass$, for $\tanb=35$, $A=0$ and a
full scan in $m$ and $M$ for the different choices of non-universal
scalar parameters \eq{nunivsc}. All the points represented fulfil the
different experimental constraints, and among them dark gray points
are those with a relic density in the range $0.1\leq\relic\,h^2\leq0.3$
and black ones correspond to those reproducing the WMAP result.
Those points excluded due to the presence of UFB minima are
shown explicitly with circles.

\begin{figure}
  \hspace*{-1.5cm}\epsfig{file=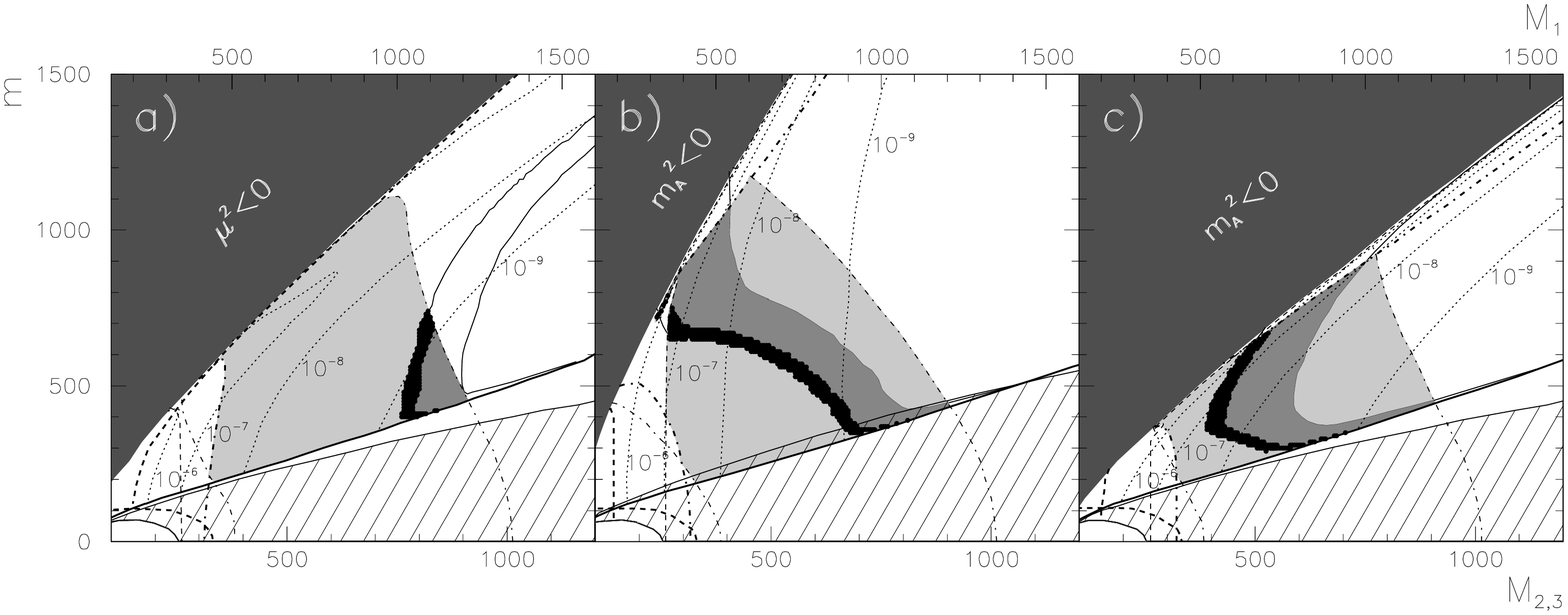,width=18cm}
  \captions{Scalar neutralino-proton cross section
    $\sigma_{\tilde{\chi}_1^0-p}$ 
    in the parameter space
    $(m,M_i)$  for
    $\delta'_{2,3}=-0.25$ and the three choices for
    non-universal scalars \eq{nunivsc}
    in a case with $\tan\beta=35$
    and $A=0$.
    The dotted curves are contours of
    $\sigma_{\tilde{\chi}_1^0-p}$. The region to the left of the 
    dashed line is excluded by the lower bound 
    on the Higgs mass.
    The region to the left of the double dashed line
    is excluded by the lower bound on the chargino mass
    $m_{\tilde\chi_1^{\pm}}>103.5$ GeV.
    The corner in the lower left shown also by a double dashed line
    is excluded by the LEP bound on the stau mass
    $m_{\tilde{\tau}_1}>87$ GeV,
    and the white region at the bottom bounded by a solid line is excluded
    because $m_{\tilde{\tau}_1}^2$ becomes negative.
    The region bounded by dot-dashed lines is allowed by $g_{\mu}-2$.
    The region to the left of the 
    double dot-dashed line is excluded by $b\to s\gamma$.
    From bottom to top, the solid lines are the upper bounds of the
    areas 
    such as $m_{\tilde{\tau}_1}<m_{\tilde{\chi}_1^0}$ (double solid), 
    $\Omega_{\tilde{\chi}_1^0} h^2<0.1$ and $\Omega_{\tilde{\chi}_1^0}
    h^2<0.3$.  The light shaded area is favoured by all the
    phenomenological constraints, while the dark one fulfils in
    addition $0.1\leq \Omega_{\tilde{\chi}_1^0}h^2\leq 0.3$. The black
    region on top of this indicates the WMAP range, 
    $0.094\leq \Omega_{\tilde{\chi}_1^0}h^2\leq 0.129$.
    The ruled region is excluded because of
    the charge and colour breaking constraint UFB-3. 
    The value of $M_3$
    is represented in the lower x-axis, whereas $M_1$ is represented in
    the upper x-axis.}
  \label{mm35n}
\end{figure}

The sensitivities of present and projected dark matter experiments are
also depicted for comparison.
The small area bounded by dotted lines is allowed by the DAMA
experiment in the simple case of an isothermal spherical
halo model. The larger area also bounded by dotted lines represents
the DAMA region
when uncertainties to this simple model are taken into account.
The (upper) areas bounded by solid lines are
excluded by CDMS and EDELWEISS.
Finally, the dashed lines represent the
sensitivities of the projected GEDEON, SOUDAN, and GENIUS
experiments\footnote{It is necessary to emphasize at
  this point that  
  the analysis including uncertainties on the isothermal spherical
  halo model has only been performed for DAMA, but not for the
  other detectors. This is a complicated issue (see e.g. \cite{copi}),
  and therefore a proper comparison (and
  determination of the real extent of the
  allowed region) is currently unavailable.}. 

The results for the neutralino-nucleon cross section are similar to
those with only scalar non-universalities (compare them with 
Figs.\,13, 15, 17 of
\cite{cggm03-1}).
In particular, regions of
the parameter space fulfilling all the constraints and with a cross
section close  to the detection range appear for moderate values of
$\tan\beta$, entering the DAMA region for $\tanb\gsim30$.
However, these regions are
shifted towards larger $M_1$ and thus heavier neutralinos are
obtained.
For
instance, in case c)
it is possible to have neutralinos compatible with DAMA with masses
as large as $300$ GeV.

In case b) a disconnected region appears with large
values for the detection cross section, $\crosssec\gsim2\times10^{-6}$
pb. Such predictions are due to the occurrence of very light Higgses, 
$91$ GeV$<m_h\lsim105$ GeV, with $\sin^2(\alpha-\beta)<0.2$, where $\alpha$
is the mixing angle in the Higgs mass matrix. 
Higgses with these properties would have escaped detection, due to the
reduction of the $ZZh$ coupling,
and are thus in agreement with the experimental bound derived from
LEP2 
\cite{barate}.
The points that would interpolate between this region and the bulk
area present heavier Higgses, $105$ GeV$\lsim m_h<114.1$ GeV, but also
larger values of $\sin^2(\alpha-\beta)$, and are therefore excluded by
the experimental constraint. In the bulk region
$\sin^2(\alpha-\beta)\approx1$ and $m_h>114.1$ GeV, thus being
experimentally allowed\footnote{In our computation the value of
  $\sin^2(\alpha-\beta)$ 
  is calculated for all points of the parameter space in order to
  apply the appropriate bound on the mass of the lightest Higgs.}.
In the remainder of the paper we will encounter similar situations,
when the choice b) in \eq{nunivsc} for scalar non-universalities is
taken. 

The corresponding $(m,M_i)$ parameter space is
represented in \fig{mm35n}
for each case,
displaying the effect of the different constraints and evidencing the
increase in $M_1$.
Because the allowed range in $M_{2,3}$ (represented in the lower
x-axis) is practically
the same as with just non-universal scalars, the values
of $M_1$ are larger. In this particular case, $M_1\gsim350$ GeV is
required.
Note that due to the reduction in the value of $\higgsu$ achieved both
through the gaugino and scalar non-universalities the regions in the
parameter space excluded due to UFB minima are smaller than in the
universal case and are
not relevant for most of the points reproducing the WMAP
result.
Obviously, this is more patent in cases a) and c) due to the choice
$\delta_2>0$ in \eq{nunivsc}, whereas in case b) the coannihilation
region with the lightest stau would still be excluded.
A small disconnected area in case b) is allowed by the experimental
constraint on the Higgs mass. It can be found close to the region
where $m_A^2$ becomes negative, and
corresponds to those points in Fig.\,\ref{cross35n}b
with larger cross section which were discussed above. 
Note also that in this area the
CP-odd Higgs is also very light, $m_A\lsim100$ GeV.

As commented above, a consequence of the decrease in $M_3/M_1$ and
$M_2/M_1$ is the reduction in the value of the relic density. This may
be problematic, since the choices of non-universal scalars
\eq{nunivsc} 
already lead to a similar decrease, specially those where the Higgsino
components of $\neut$ increase. This is the case of example a), for
those points close
to the upper-left corner (which are excluded 
because $\mu^2<0$).
Also for this choice of $\tan\beta$ in example a)
the value of $m_A$ is very close to $2\neumass$ in most of the
parameter space, thus boosting 
the annihilation through the resonant $s$-channel and implying that
most of the points allowed by experimental constraints have
a too low relic density. Only a few points with a mostly bino
composition and a high neutralino mass, $\neumass\gsim450$ GeV  are
left in this case with $\crosssec\lsim10^{-9}$ pb.
In examples b) and c), 
regions which survive once the WMAP constraint is
applied are found for lighter neutralinos. 
Due to the more effective reduction of $m_A$ the resonant
neutralino annihilation takes place for smaller values of $\tan\beta$
in these cases.

\begin{figure}
  \hspace*{-1.5cm}\epsfig{file=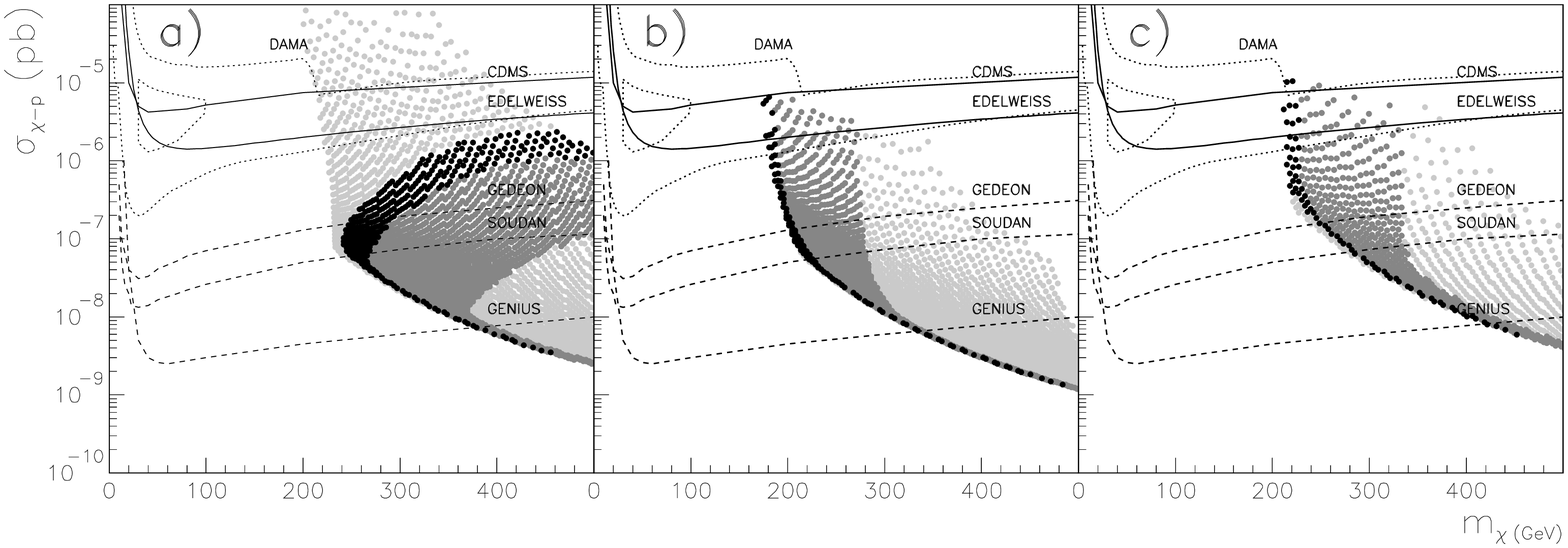,width=18cm}
  \captions{
    The same as in \fig{cross35n} but for $\tan\beta=50$.}
  \label{cross50n}

\vspace*{1cm}

  \hspace*{-1.5cm}\epsfig{file=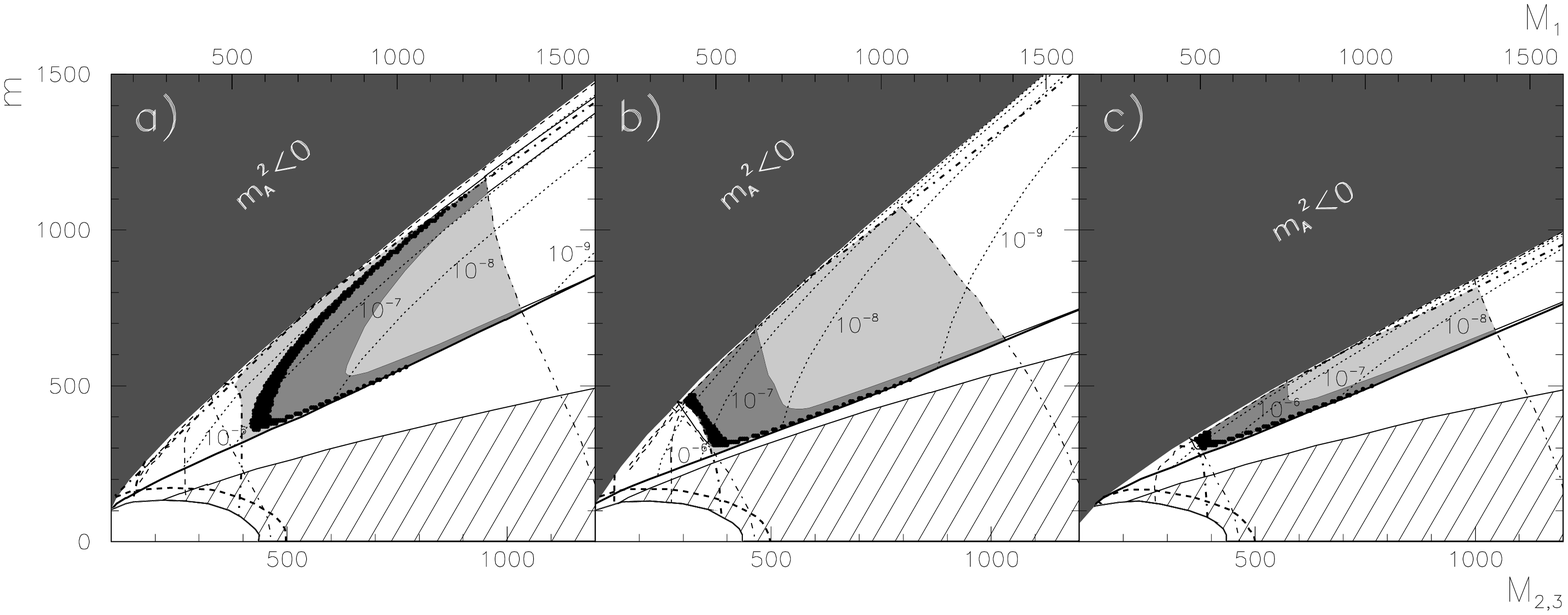,width=18cm}
  \captions{
    The same as in \fig{mm35n} but for $\tan\beta=50$.}
  \label{mm50n}
\end{figure}

Increasing the value of $\tan\beta$ leads to the well known 
enhancement of $\crosssec$. 
An example with
$\tan\beta=50$ is represented in \fig{cross50n}, where points close to
the sensitivities of present experiments and compatible with all the
constraints appear for all three cases a), b) and c). Due to the
reduction in $m_A$
in case a) 
all the points are
already beyond the resonance, some having the correct relic density,
and the regions leading to pure Higgsino-like
neutralinos are now excluded due to the occurrence of a tachyonic
CP-odd Higgs. 
In all the examples 
the points with a higher $\crosssec$
correspond to those having $m_A$ close to its experimental lower
limit.
The effect of the different constraints on the $(m,M_i)$ parameter
space are explicitly shown in \fig{mm50n}, where we can see how due
the further reduction in the value of $m_A$ the
allowed regions correspond to narrower ranges of $m$ and $M_i$. Also,
the UFB constraints are less restrictive and now they do not exclude
the coannihilation tail in case b).

Concerning variations in $M_2$, as we have already mentioned, 
if it is further decreased
below a critical value ($\delta'_2\lsim-0.5$) we
eventually end up with a lightest neutralino which is mainly
wino. Although such a change in the neutralino composition
highly enhances 
the cross section, the relic density
decreases and the WMAP constraint is no longer fulfilled.

\begin{figure}
  \hspace*{-1.5cm}\epsfig{file=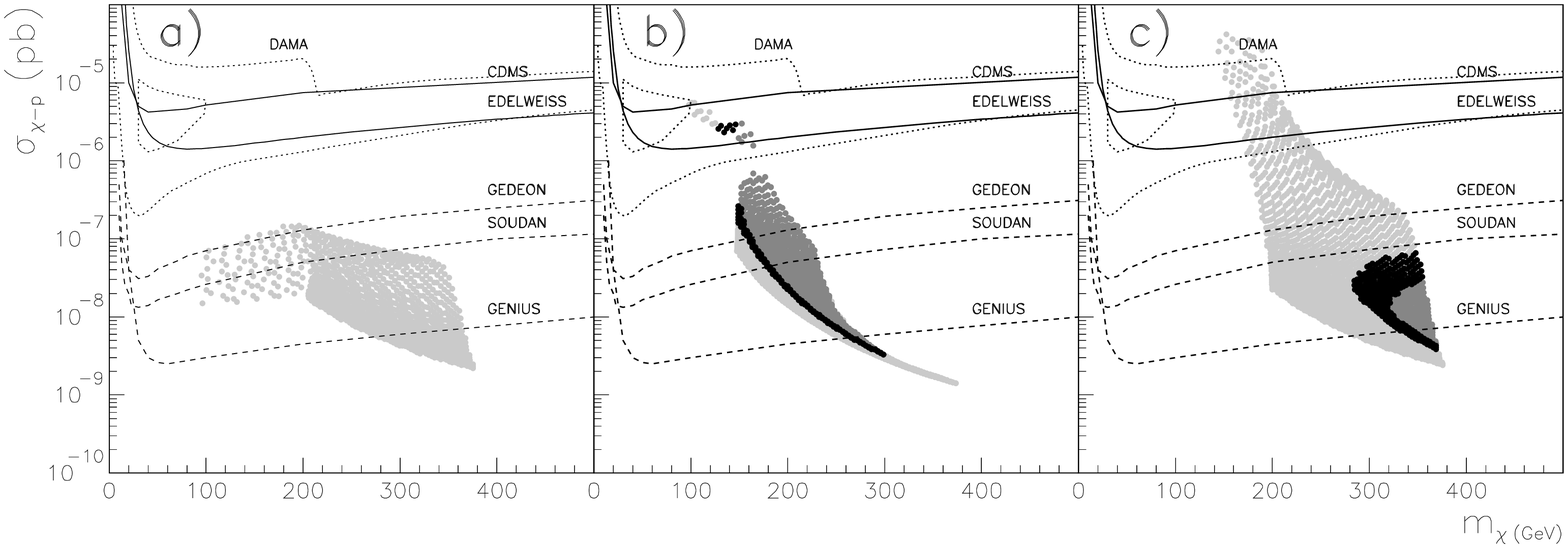,width=18cm}
  \captions{The same as in \fig{cross35n} but for 
    $\delta'_2=0.25$ and $\delta'_{3}=-0.25$}
  \label{cross35o}

  \vspace*{1cm}

  \hspace*{-1.5cm}\epsfig{file=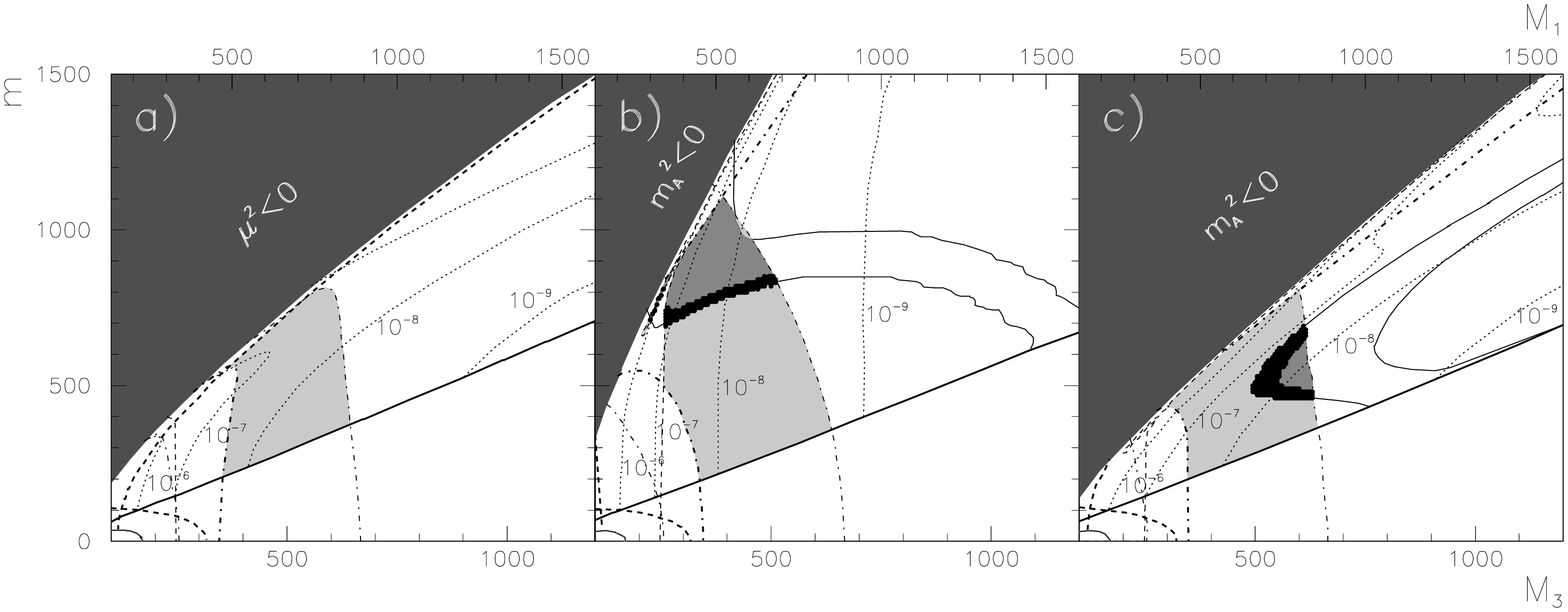,width=18cm}
  \captions{The same as in \fig{mm35n} but for
    $\delta'_2=0.25$ and $\delta'_{3}=-0.25$}
  \label{mm35o}
\end{figure}

If the value of $M_2$ 
is increased with respect to $M_1$
(thus having $M_2>M_1>M_3$ at the GUT
scale) a reduction of $\asusy$ is obtained, which
sets a stronger upper constraint for both $M$ and $m$. 
To illustrate this, we have represented in \fig{cross35o} an example
with $\delta'_2=0.25$ and $\delta'_{3}=-0.25$. Although the change in
the theoretical predictions for the neutralino-nucleon cross section
is very subtle,
the effect of the stronger $g-2$ constraint can have important
consequences.
This reduction in the parameter space can be seen in
\fig{mm35o}, where the parameter space $(m,M_i)$ is represented. In
particular, in
order to satisfy the lower limit on $\asusy$ we need to have
$M_1\lsim900$ GeV. Because of this, the regions with the correct relic
density are much smaller, as in case c), and can even be excluded.
Note also that since the value of $\higgsu$ is further 
increased no regions
in the parameter space are excluded due to the occurrence of UFB
minima in these examples.
Since once more the resonant annihilation of neutralinos is very
efficient in example a), the relic density is too low in those points
of the parameter space which fulfil all the experimental constraints
($\relic\,h^2\lesssim0.045$).

Further decreasing $\delta'_3$ leads to larger values of $M_1$ and
thus heavier neutralinos can be obtained. At the same time the $\mu$
term slightly decreases and eventually it can be of the same order or
even smaller than $M_1$ and thus Higgsino-like neutralinos
appear, which might have a large value for the cross section. 
However this possibility is very limited. On the one hand,
$M_2$ cannot be decreased beyond ${M_1}/{2}$ in order not to run
into the problems of a wino neutralino. On the other hand, if $M_2\gg
M_3$ the lower constraint on $\asusy$ (which sets an upper bound for
$M$) and the constraints on the Higgs
mass and \bsg\ (which set a lower bound for $M$) 
may not be simultaneously fulfilled.
Furthermore, the relic density of Higgsino-like neutralinos is
typically very low\footnote{A 
  similar scenario, with just non-universal gauginos resulting from
  the $n=200$ representation of $SU(5)$ and leading to Higgsino dark
  matter was studied in Ref.~\cite{Roy2}, where it was shown that
  their low relic density is below the astrophysical constraint.} 
\cite{mizutahiggs},
and consistency with the WMAP result
is not always obtained.
Obviously, Higgsino dark matter will be more easily obtained for those
choices of non-universal scalars \eq{nunivsc}
with $\delta_2>0$ (examples a) and c)),
since they lead to a very effective decrease in the $\mu$
parameter. It is in these cases where the problems associated to
Higgsino-like neutralinos are more patent\footnote{Choosing 
  $\delta_2<0$ (thus having more negative values for $\higgsu$) 
  leads to an increase of the $\mu$ parameter and
  can help restoring the gaugino character of the lightest
  neutralino. Heavier bino-like neutralinos
  satisfying the astrophysical constraint on the relic density can
  therefore be obtained, but
  the neutralino-nucleon cross section has a significant
  decrease, due to both the increase in $\mu$ and in $m_A$. 
  Also the lightest neutralino is not the LSP in larger regions
  in the parameter space.}.

\begin{figure}
  \hspace*{-1.5cm}\epsfig{file=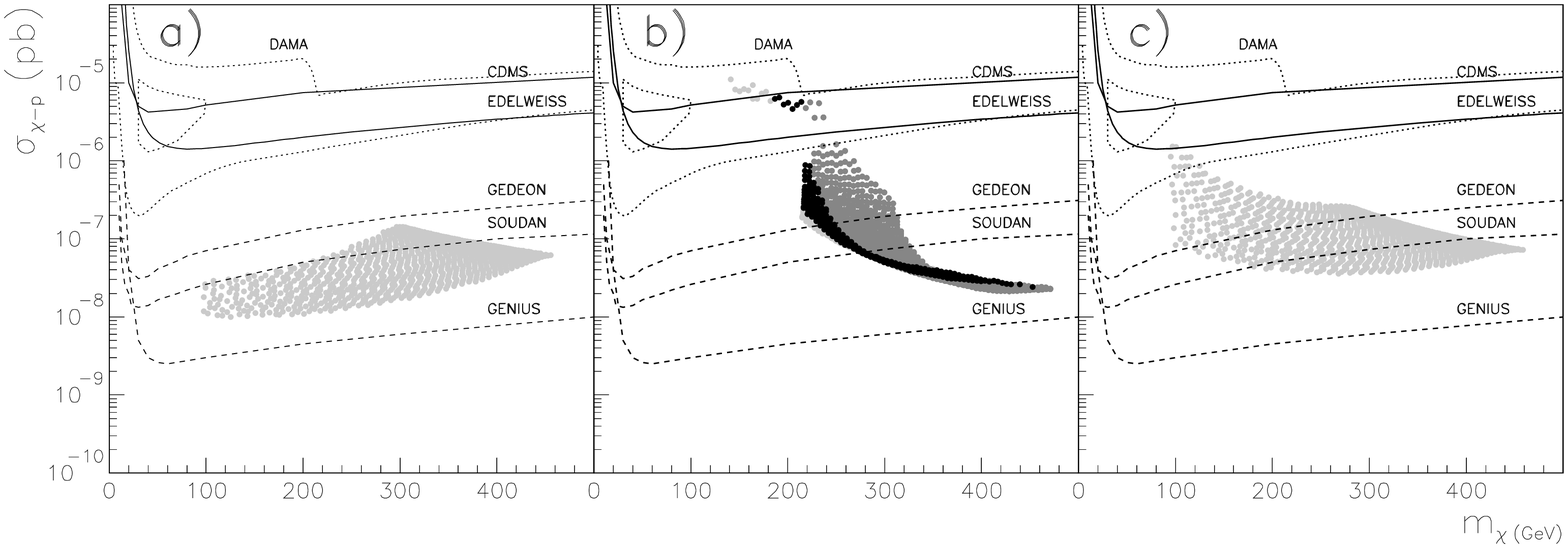,width=18cm}
  \captions{The same as in \fig{cross35n} but for $\delta'_{2}=0$ and
    $\delta'_{3}=-0.5$}
  \label{cross35z}

  \vspace*{1cm}

  \hspace*{-1.5cm}\epsfig{file=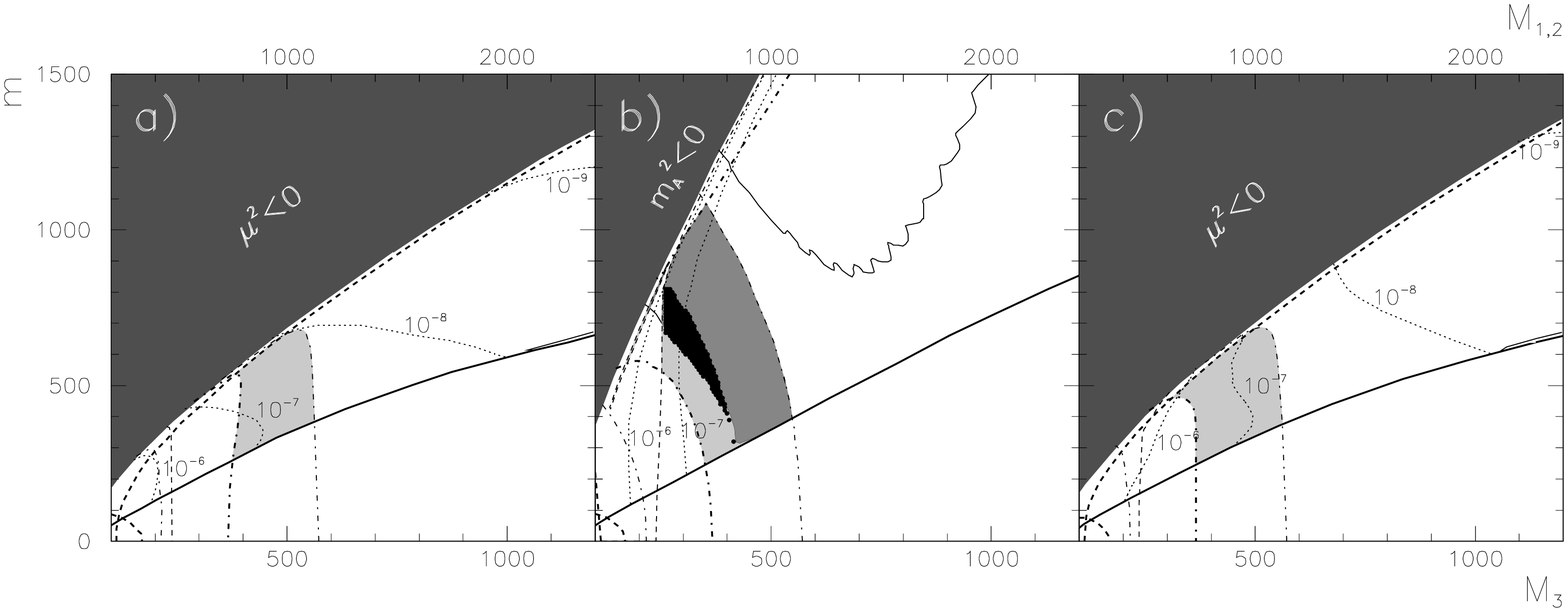,width=18cm}
  \captions{The same as in \fig{mm35n} but for $\delta'_{2}=0$ and
    $\delta'_{3}=-0.5$}
  \label{mm35z}
\end{figure}

The predicted $\crosssec$ for an example where
$\delta'_{2}=0$ and $\delta'_{3}=-0.5$ and the three choices for
non-universal Higgses \eq{nunivsc} have been taken is illustrated in
\fig{cross35z}, 
with the corresponding $(m,M_i)$ parameter space in
\fig{mm35z}. 
As a consequence of the
decrease in $M_3$ the constraints on the Higgs mass and \bsg\ are only
fulfilled for large values of $M_{1,2}$, which are almost
comparable to the constraint due to the lower bound on $\asusy$ and
the parameter space is very reduced. 
Due to the further decrease in $\mu$, those regions excluded for
having $\mu^2<0$ are now slightly larger, as is the case of example
a). Note in this sense the upper region in case c) which is now also
excluded for this reason ($\mu^2$ now becomes negative before
$m_A^2$).
The Higgsino composition of the lightest neutralino is very important
in both a) and c), with $0.3\lsim N_{13}^2+N_{14}^2\lsim1$, leading to
light neutralinos ($\neumass\gsim100$ GeV) but with
very low values for the relic density, $\relic\,h^2\lsim0.06$ and
$\relic\,h^2\lsim0.07$, respectively.
On the contrary in case b) the neutralinos still continue being mostly
binos ($N_{13}^2+N_{14}^2\lsim0.13$)
and
points fulfilling WMAP with $\neumass\gsim200$ GeV and a high cross
section are
found. 
Note
that 
in these examples 
the UFB constraints are also satisfied in the whole
parameter space due to the less negative values of $\higgsu$.

Let us finally remark that all of the above results 
were obtained for $A=0$. 
Departures from this value can alter the results for the
neutralino-nucleon cross section and the relic
density. 
In particular, for positive values of the trilinear term, the negative
contributions in the RGE of the Higgs parameters due to $A^2$ terms
are less important and thus both $\higgsd$ and $\higgsu$ increase.
This entails a slight increase in $\crosssec$ and a decrease in
$\relic$, as well as a reduction in the region restricted by the UFB
constraints, with opposite effects for negative values of $A$.
In some cases, as for instance, in example a) in Figs.\,\ref{cross35n}
and \ref{cross35o} the shift in $m_A$ due to variations in the
trilinear parameter is enough to avoid 
the resonant neutralino annihilation and regain the correct $\relic$
in parts of the parameter space.

\subsection{Increase in $M_3/M_1$}

Let us now analyse the other possibility, namely increasing the value
of $M_3$ with respect to $M_1$, which can be done with
$\delta'_3>0$ in \eq{gauginospara}.
In this case, 
the constraint on the Higgs mass and on \bsg\ 
will be satisfied for smaller
values of $M$, and therefore the effective value of $M_1$ can be
smaller than in the universal case.
Thus lighter neutralinos can be obtained.

\begin{figure}
  \hspace*{-1.5cm}\epsfig{file=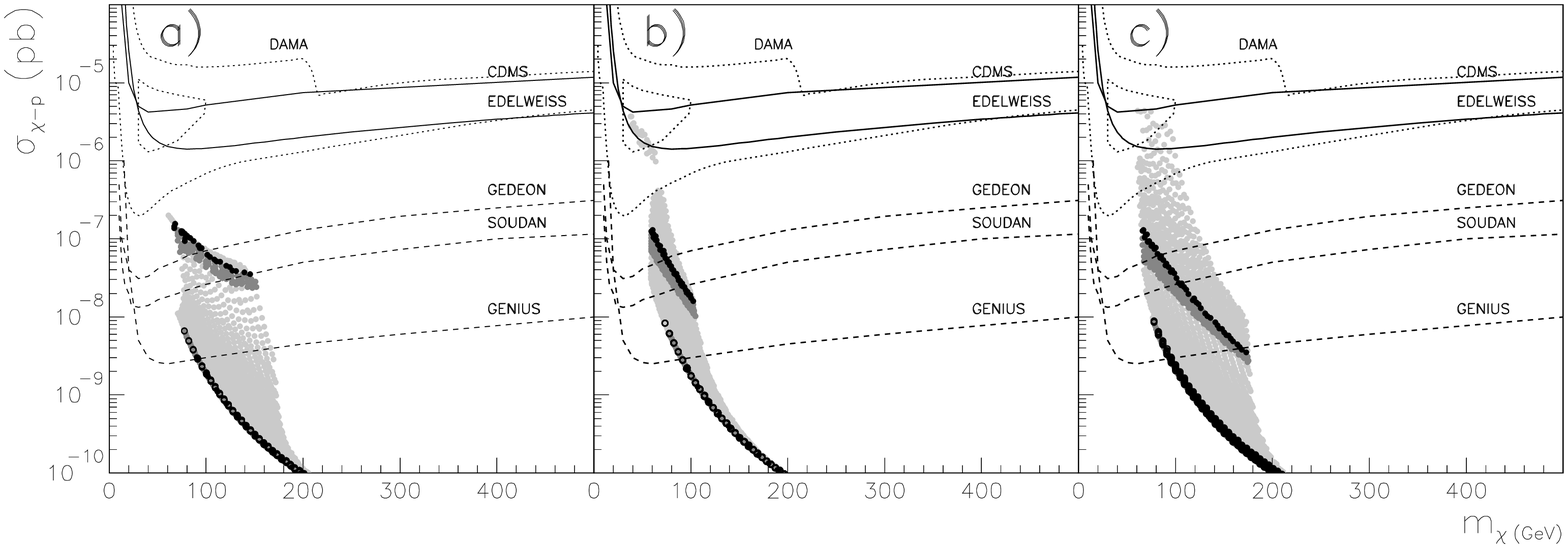,width=18cm}
  \captions{The same as in \fig{cross35n} but for 
    $\delta'_{2,3}=1$}
  \label{cross35w}

  \vspace*{1cm}

  \hspace*{-1.5cm}\epsfig{file=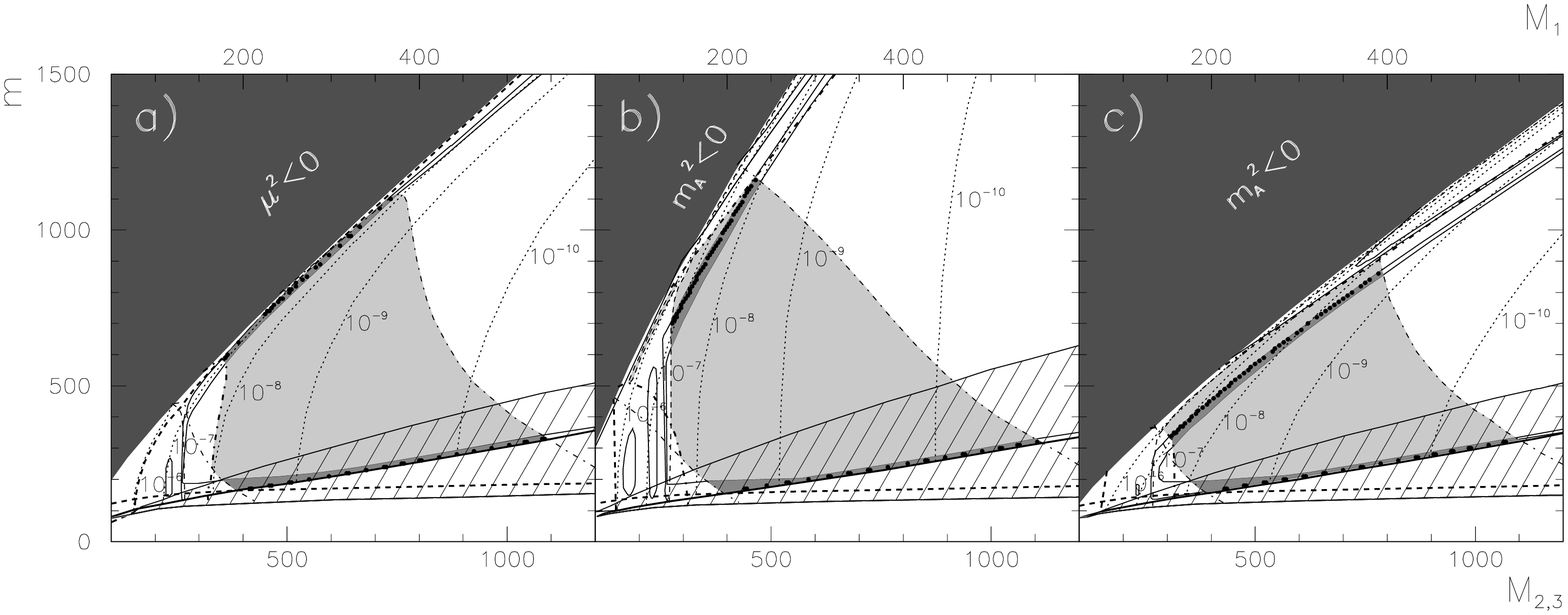,width=18cm}
  \captions{The same as in \fig{mm35n} but for
    $\delta'_{2,3}=1$  
    }
  \label{mm35w}
\end{figure}

\begin{figure}
  \hspace*{-1.5cm}\epsfig{file=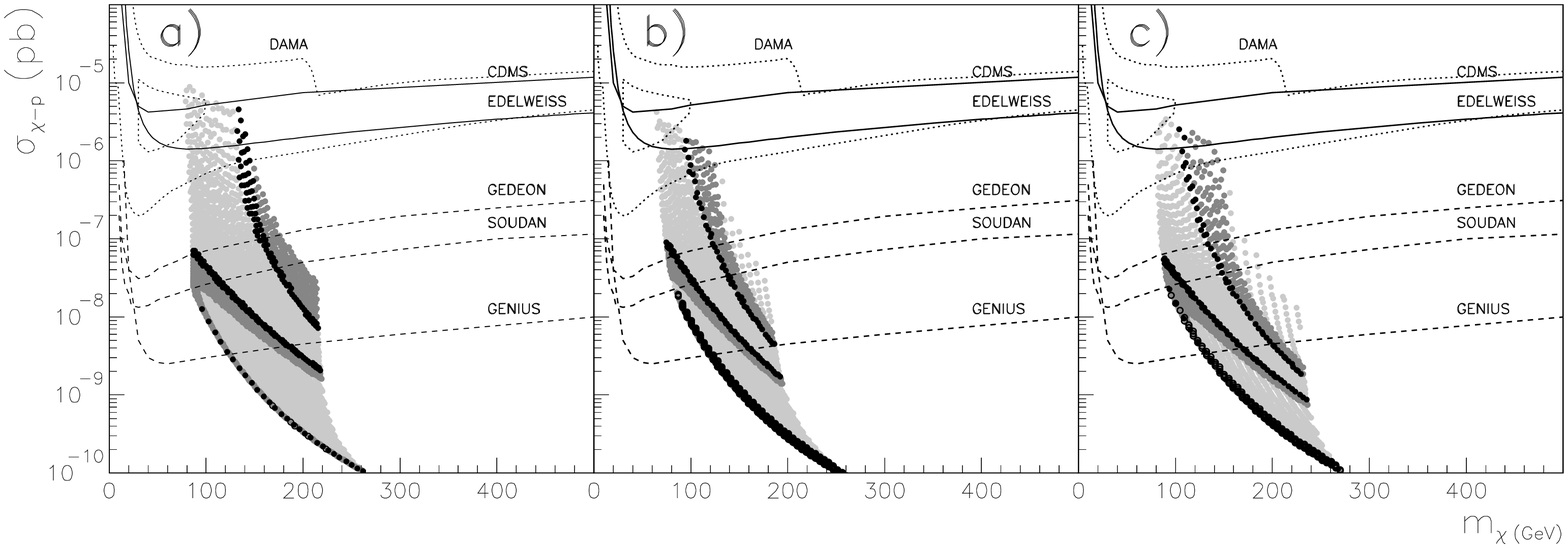,width=18cm}
  \captions{The same as in \fig{cross35w} but for $\tan\beta=50$}
  \label{cross50w}

  \vspace*{1cm}

  \hspace*{-1.5cm}\epsfig{file=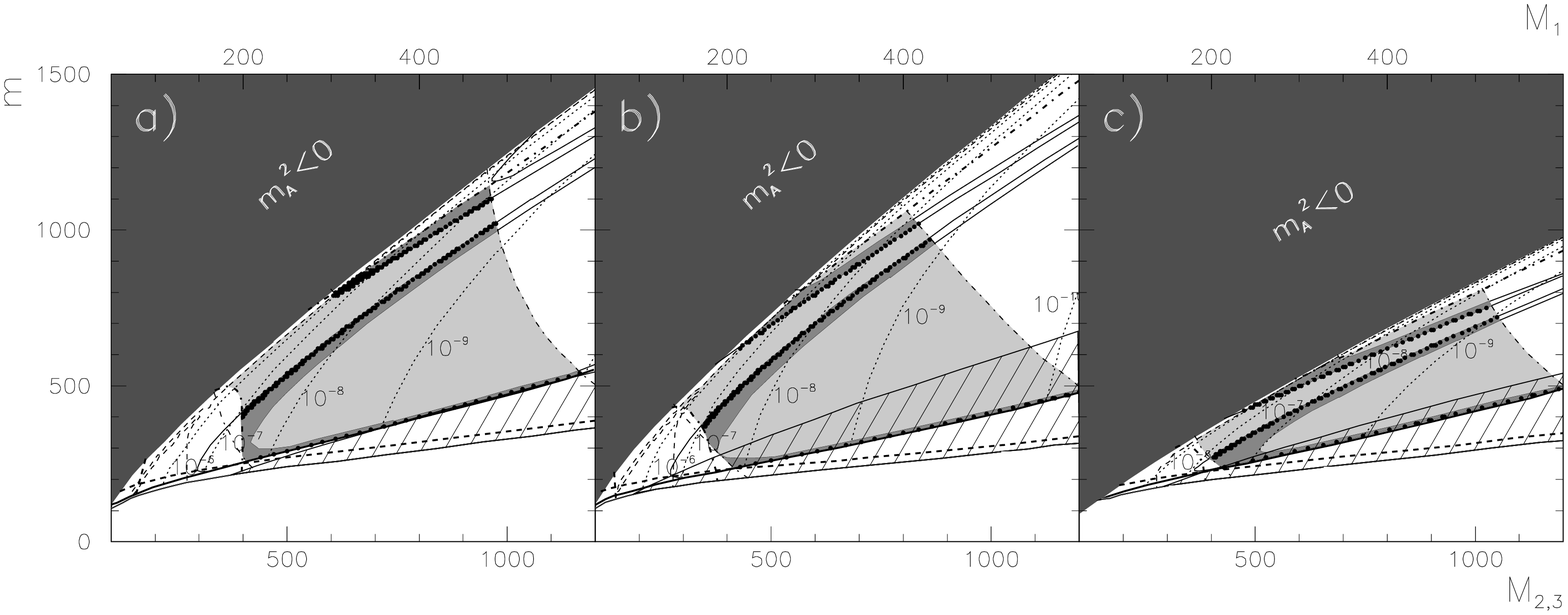,width=18cm}
  \captions{The same as in \fig{mm35w} but for $\tan\beta=50$}
  \label{mm50w}
\end{figure}

Regarding the value of $M_2$, let us begin by considering
also an increase in $M_2/M_1$ and discuss
departures from this choice later. The
structure of soft masses at the GUT scale would therefore 
be $M_3 \sim M_2 > M_1$.
The theoretical predictions for $\crosssec$  are represented in
\fig{cross35w} for
an example with $\delta'_{2,3}=1$, $\tan\beta=35$ and $A=0$ and the
three choices of
Higgs non-universalities of \eq{nunivsc}.
As we can see, this choice of gaugino parameters
favours the appearance of light neutralinos which obviously have
a large bino component. 
The predicted cross section is only slightly smaller than in the
cases with just non-universal scalars, so these neutralinos can still
be close to the sensitivities of dark matter experiments. In
particular, neutralinos with $\crosssec\gsim10^{-7}$
pb can be obtained with $\neumass\sim60$ GeV for the three cases a),
b) and c).

This
effective reduction of the value of $M_1$ is clearly 
manifest in the plots representing the corresponding 
$(m,M_i)$ parameter space in
\fig{mm35w}. Because 
of the increase of $M_3$ the regions excluded due to the constrains on
\bsg\ and the Higgs mass now occur for $M_1\lsim 180$ GeV.
Concerning the UFB constraints, they are more restrictive than in the
examples of the previous section, due to the decrease in $\higgsu$,
and exclude larger regions in the parameter space. 
In particular, those regions having the
correct relic density due to coannihilations with the NLSP are ruled
out for this reason. However, points where the reduction in the
relic density is due to a decrease in $m_A$ are still allowed, giving
rise to narrow allowed regions.

Increasing the value of $\tan\beta$ larger cross sections can be
obtained and the UFB bounds become
less stringent as a consequence of the
increase in $\higgsu$. For instance, if
$\tan\beta=50$ is taken in the former example, points satisfying all
the constraints and entering the DAMA region can be obtained. 
The
predictions for $\crosssec$ in this case are
shown in \fig{cross50w} and the corresponding $(m,M_i)$ parameter
space is represented in \fig{mm50w}. In cases a) and c), where the
increase in $\higgsu$ is more effective, the UFB constraints are
weaker and for instance in case a) they do not exclude completely 
the coannihilation tail with the lightest stau.

\begin{figure}
  \hspace*{-1.5cm}\epsfig{file=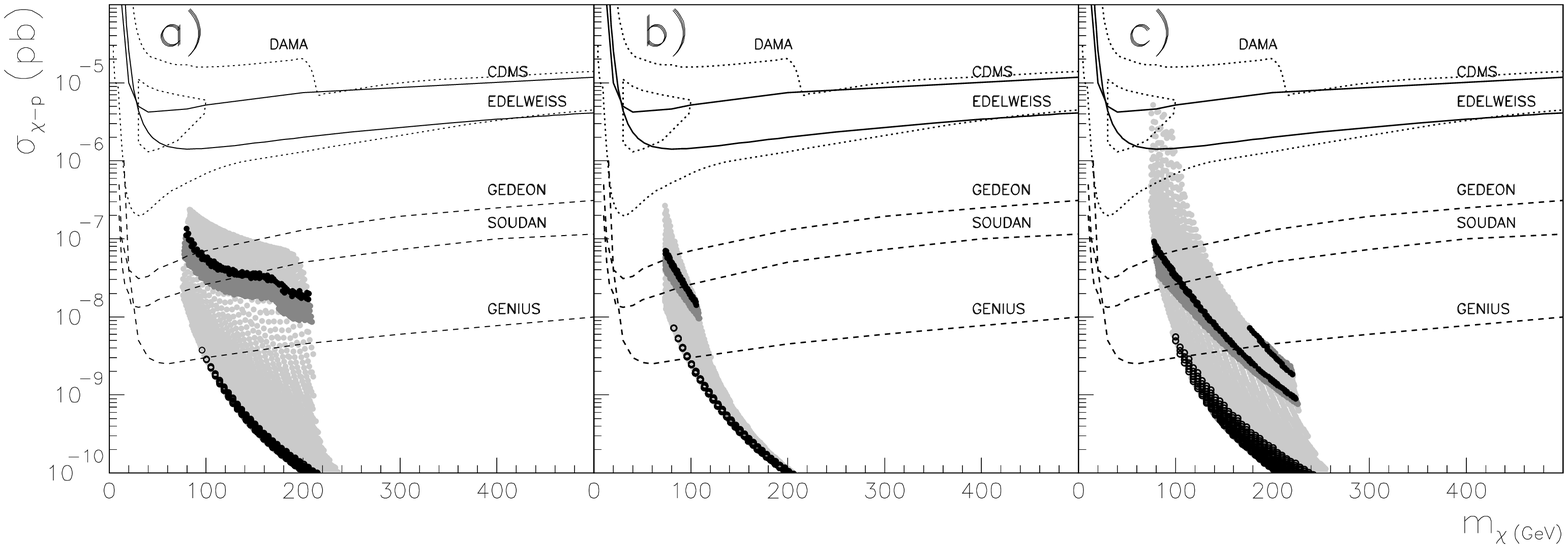,width=18cm}
  \captions{The same as in \fig{cross35n} but for 
    $\delta'_{2}=-0.25$ and $\delta'_{3}=1$}
  \label{cross35u}

  \vspace*{1cm}

  \hspace*{-1.5cm}\epsfig{file=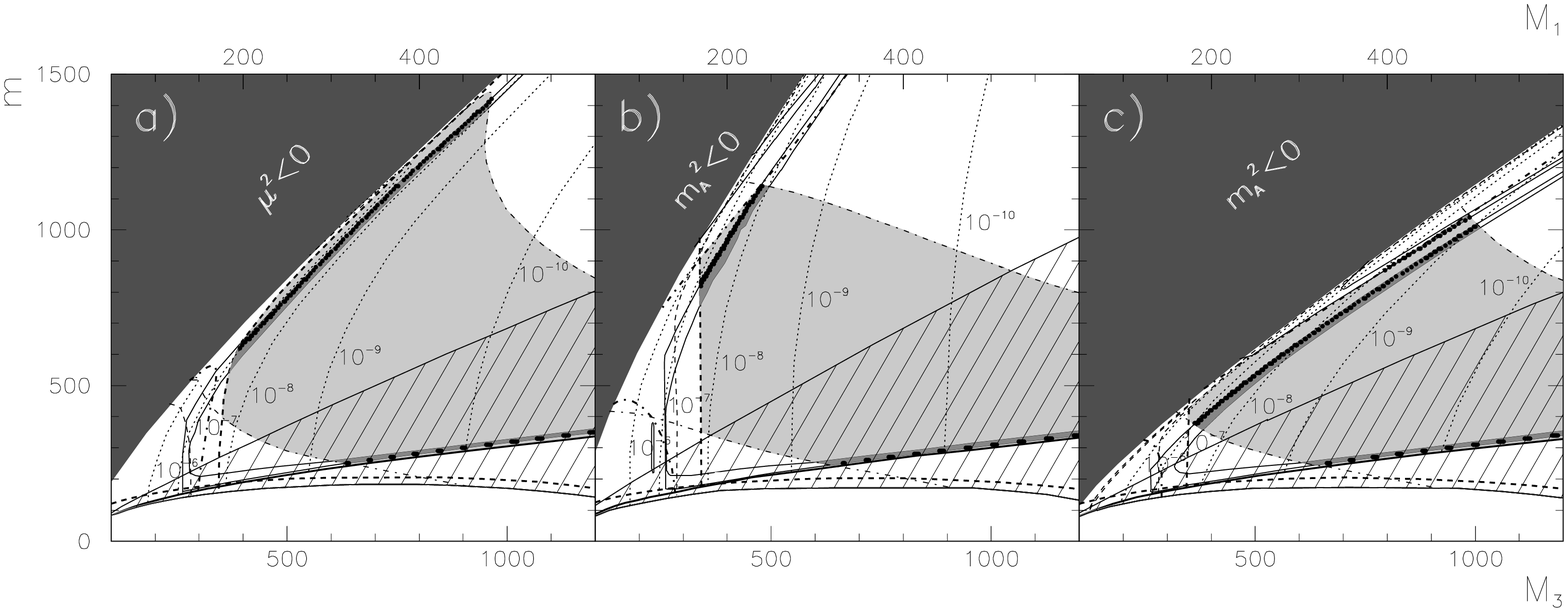,width=18cm}
  \captions{The same as in \fig{mm35n} but for 
    $\delta'_{2}=-0.25$ and $\delta'_{3}=1$}
  \label{mm35u}
\end{figure}

Let us now comment on the possibility of decreasing $M_2/M_1$. Once
more, in order not to have the problems
associated to a neutralino with a large wino composition we will
restrict this decrease to $M_2/M_1>0.5$ ($\delta'_2>-0.5$). 
The structure of gaugino masses at the GUT scale 
in this case would therefore be
$M_3 > M_1 \gsim M_2$. 
The theoretical predictions for $\crosssec$ for an example with
$\delta'_2=-0.25$ and $\delta'_3=1$ are represented in \fig{cross35u},
showing 
again very subtle variations with respect to the $\delta'_2=\delta'_3$
case.
The corresponding $(m,M_i)$ parameter space, with the
experimental and astrophysical 
bounds, is shown in \fig{mm35u}. There we can see
the shift of the regions excluded by $\asusy$ towards 
higher values of $M$, as
well as the effect that the lightest bound on the chargino
mass has in restricting the parameter space. The chargino bound 
can become more
important than the constraints due the lightest Higgs mass and the
\bsg. For instance, in cases a) and b) it restricts the allowed area
to $M_1\gsim175$ GeV for this 
choice of $\tan\beta$, thus setting a stringent limit on the
appearance of light neutralinos\footnote{
  Note that we are using here $\charmass>103.5$ GeV as the lower bound
  on the chargino mass \cite{chargino}, 
  which is in fact only valid in the case of gaugino
  unification. This bound can be relaxed to $\charmass\gsim90$ GeV in
  non-universal scenarios (see e.g. the discussion in \cite{heister}). 
  In such a case we would obtain a slightly
  larger allowed
  area in the parameter space, since now $M_1\gsim155$ GeV,
  and therefore slightly lighter neutralinos. 
  The increase in the predictions for $\crosssec$ in the area
  allowed by WMAP is, however, insignificant.
  }.

\begin{figure}
  \hspace*{-1.5cm}\epsfig{file=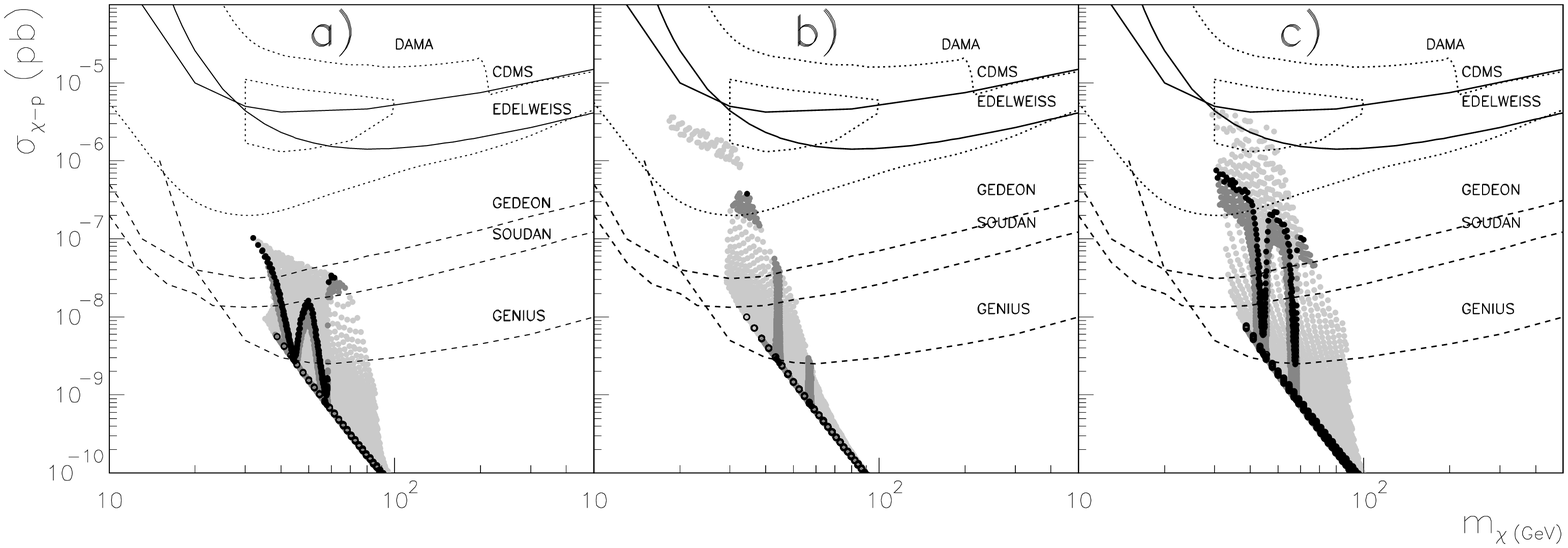,width=18cm}
  \captions{The same as in \fig{cross35n} but for 
    $\delta'_{2,3}=3$.}
  \label{cross35s}

  \vspace*{1cm}

  \hspace*{-1.5cm}\epsfig{file=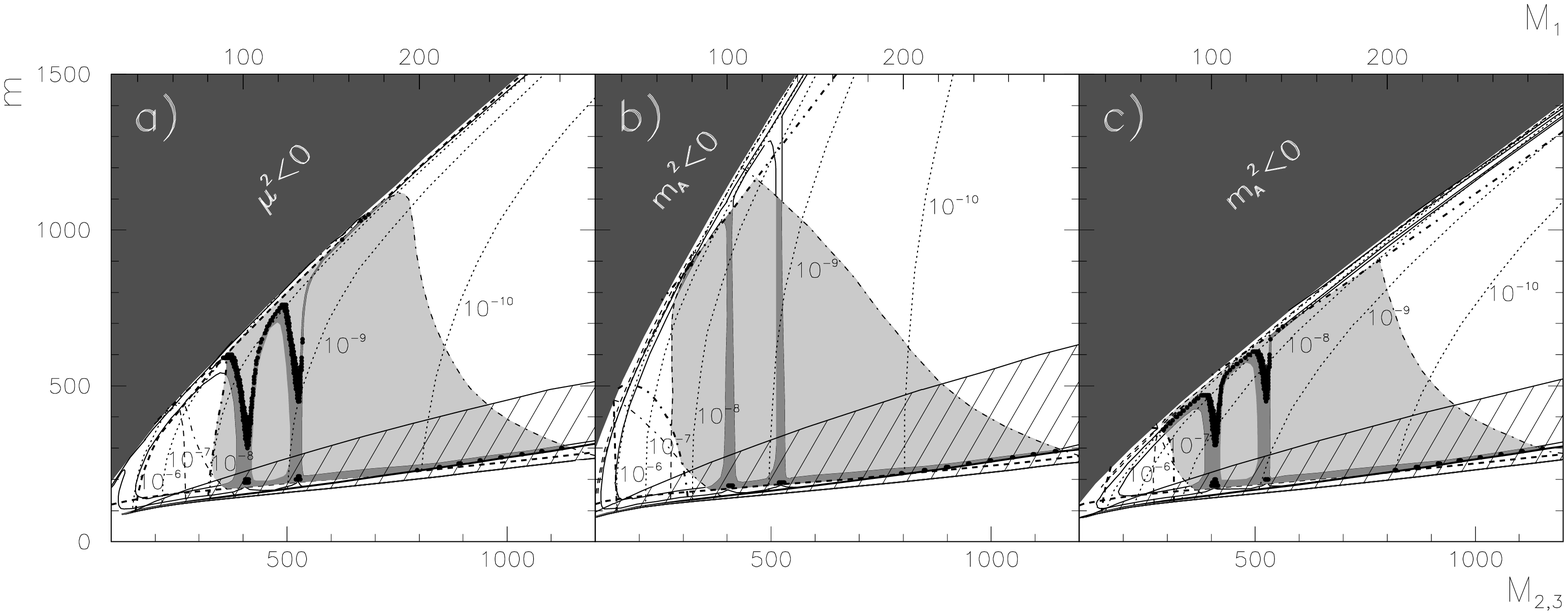,width=18cm}
  \captions{The same as in \fig{mm35n} but for 
    $\delta'_{2,3}=3$.}
  \label{mm35s}
\end{figure}

The value of $M_1$ can be further decreased if larger values of
$\delta_3$ are used, thus leading to even lighter neutralinos. 
In order to illustrate this possibility an example with
$\delta'_{2,3}=3$ is 
represented in \fig{cross35s}. Very light neutralinos can
appear within the DAMA sensitivity range.
Usually in these cases the typical values for the relic density
are too large and are therefore not consistent with the WMAP result. 
As usual a very effective decrease can be achieved when
the $h$ and $Z$-poles are crossed at $\neumass=m_h/2$ and $M_Z/2$,
respectively, giving rise to a very effective neutralino
annihilation through the corresponding $s$-channels. This 
is evidenced by the narrow
chimneys in the cosmologically preferred regions.
However, there is now a new interesting possibility. Because of the
very efficient decrease in the CP-odd Higgs mass, annihilation of very
light neutralinos can be boosted and thus the correct relic density
obtained. This happens in our example for case c), allowing the
existence of 
neutralinos with $\neumass\sim30$ GeV which are compatible with the
DAMA region.

As we have already commented in Sections~\ref{sec_nunivsc} and
\ref{sec_nunivg}, 
such light neutralinos cannot be
obtained in SUGRA theories with non-universalities in
just the scalar or gaugino sector.
Let us therefore study this possibility in more detail
within the framework of these more general SUGRA theories.

\subsubsection{Very light neutralinos}
\label{sec_verylight}

The flexibility due to non-universal gauginos
was recently exploited in
\cite{bbpr-hooper,bottinog,lightneut} in order to
calculate a lower bound for the lightest neutralino in the effMSSM,
where the parameters are defined directly at the electroweak scale
(for previous works see \cite{dreesroszgabutti}). 
The relic density of very light neutralinos ($\neumass<M_Z/2$) is a
decreasing function of $\neumass$ and therefore  a lower bound on
$\neumass$ can be extracted from the upper bound
on $\relic$. 
It was shown \cite{bottinog,lightneut}
that although 
the relic density of such
light neutralinos
usually exceeds the upper bound, a significant
reduction can be obtained when the mass of the CP-odd Higgs is small
($m_A\lsim200$ GeV) and for large values of $\tan\beta$.
Under these conditions a lower limit
$\neumass\gsim6$ GeV was extracted \cite{bottinog}, which was also
found to be consistent with the experimental constraints
\cite{lightneut}.

One of the requirements for the appearance of such very low
neutralinos is to have $M_1\ll\mu, M_2$ at low energy (thus having
almost pure binos). This can be achieved with adequate choices of
gaugino non-universalities, in particular with $\delta'_{2,3}\gg1$. 
However, as mentioned above,
without a very effective reduction of $m_A$, the relic density would
be too large, and therefore inconsistent with observations. 
Here the presence of non-universal scalars is
crucial. In particular, non-universalities as the ones we have
described in the Higgs sector in
\eq{nunivsc} provide a very effective way of lowering $m_A$ and are
thus optimal for this purpose.

More specifically, it is in case b) and especially c) where the
reduction in 
$m_A$ is more effective (not being so constrained by regions with
$\mu^2<0$) and for this reason very light neutralinos 
can easily appear. We have already seen in a former example
how this happened
for case c) with $\tan\beta=35$ and $\delta'_{2,3}=3$ (see
\fig{cross35s}). On the
contrary, in case a) higher values of $\tan\beta$ are required in
order to further reduce the value of $m_A$. 
We have checked explicitly that 
$\tan\beta\gsim33$ is sufficient to obtain $\neumass<M_Z/2$ in cases
b) and c), whereas $\tan\beta\gsim45$ is necessary in case a). In all
the three cases $\delta'_{2,3}\gsim3$ leads to these results.

\begin{figure}
  \hspace*{-1.5cm}\epsfig{file=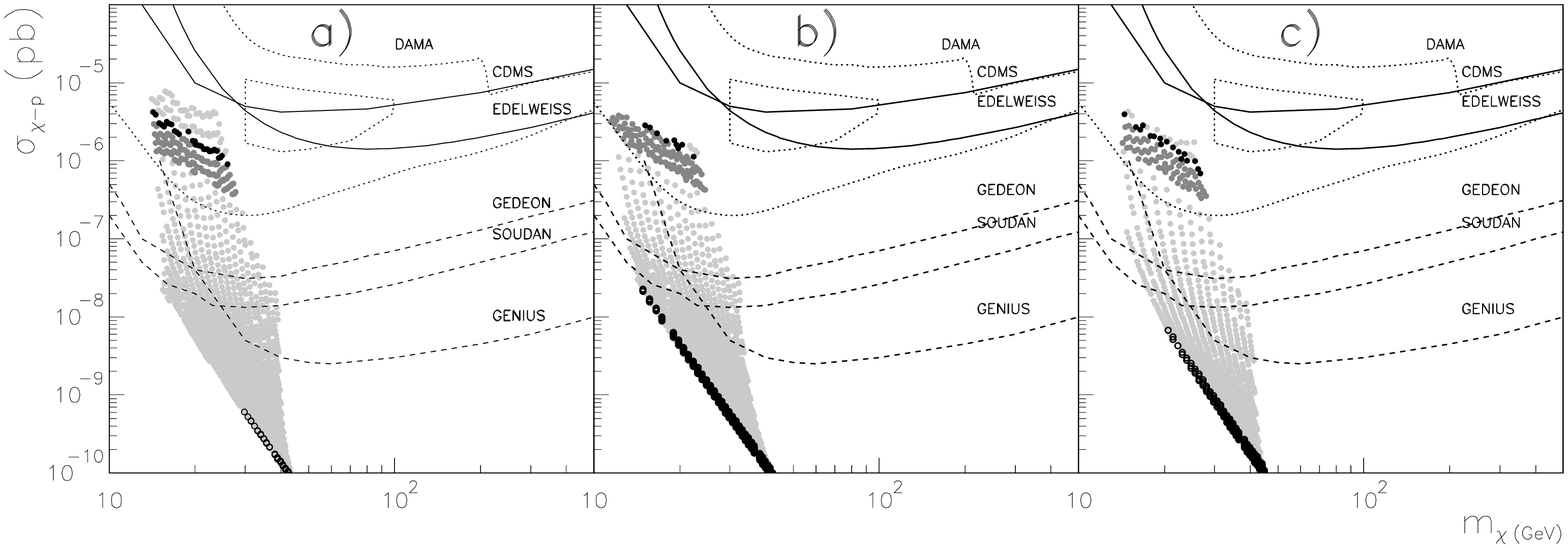,width=18cm}
  \captions{The same as in \fig{cross35n} but for 
    $\delta'_{2,3}=10$ and  $\tan\beta=50$.}
  \label{cross50q}

  \vspace*{1cm}

  \hspace*{-1.5cm}\epsfig{file=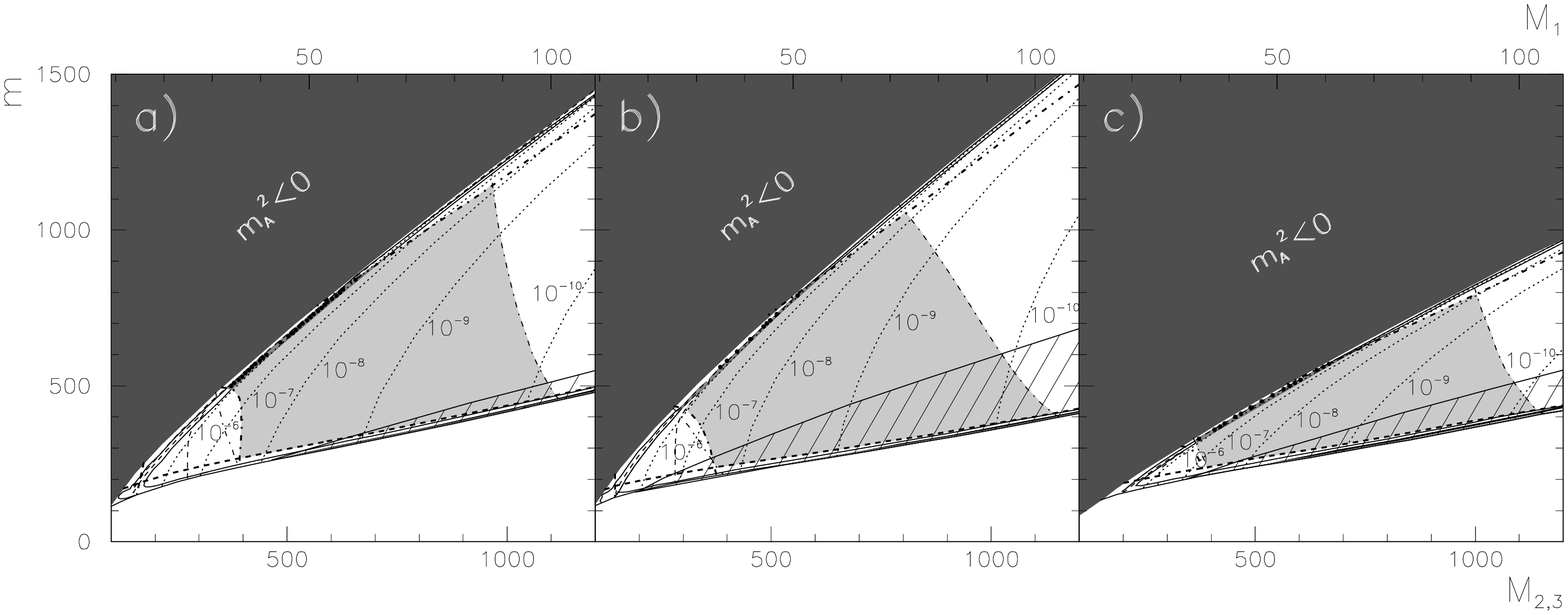,width=18cm}
  \captions{The same as in \fig{mm35n} but for
    $\delta'_{2,3}=10$ and  $\tan\beta=50$.}
  \label{mm50q}
\end{figure}

Obviously, lighter neutralinos can be obtained if $\delta'_{2,3}$ are
increased. 
Let us therefore complete our discussion 
by analysing the case $\delta'_{2,3}=10$ for the three choices of
scalar non-universalities \eq{nunivsc}. The resulting
neutralino-nucleon cross section versus the neutralino mass is
represented in \fig{cross50q} for $\tan\beta=50$ and $A=0$, together
with the sensitivities of dark matter detectors. We observe
the appearance of very light neutralinos, 
whose
cross section can be in
range of detectability of near-future experiments. In particular,
points with $\crosssec\gsim3\times10^{-6}$ pb are obtained with
$\neumass\sim 15$ GeV, in agreement with the bound derived in the
effMSSM \cite{bottinog}.
Once more the
resonances with the lightest Higgs and the $Z$ give rise to the
characteristic narrow chimneys at the corresponding values of the
neutralino mass. 

The effect of the different constraints on the
corresponding $(m,M_i)$ parameter space is represented in
\fig{mm50q}. Note that the regions giving rise to very light
neutralinos with a consistent relic density are extremely narrow.
In these points the mass of the CP-odd Higgs can be very close to its
experimental limit, $m_A\lsim100$ GeV. In fact, in these cases we are
near the ``intense coupling regime'' for the Higgs sector, where
the masses of the Higgses are almost degenerate, 
and even beyond it, thus having $m_H> m_h\sim m_A$ with
$\sin\alpha\sim-1$.
Regarding the experimental bound on the lightest Higgs in this last
case, 
note that,
since $\tan\beta\gg1$, this implies 
$\sin^2(\alpha-\beta)\ll1$ and therefore the constraint on the
lightest Higgs can be relaxed
to $m_h\gsim91$ GeV \cite{barate}.
Finally, since the neutralino mass is so small, the region
excluded due to the neutralino not being the LSP is
negligible. However, now the lower bound on the stau mass
plays an important role. In fact,
an important region of the parameter space
is excluded for having $m_{\tilde{\tau}_1}^2<0$.

In those examples with $\delta_2>0$ a reduction in the
$\mu$ parameter is more easily achieved, thus obtaining typically the
structure $M_1\ll\mu< M_2$ at low energy. However, when
$\delta_1<0$ is taken (cases b) and c)), 
the more effective decrease in $m_A$ can forbid
some of the points with very low $\mu$, thus obtaining instead
$M_1\ll M_2\lsim\mu$.
This has clear implications on the
neutralino-chargino masses and
compositions. For example, in the first case, the lightest chargino 
and the second lightest neutralino would be mainly
Higgsinos, whereas in the second case they would have a larger
wino composition.
Also in case c), where both $\delta_2>0$ and $\delta_1<0$ are taken, the
resulting allowed values for the common scalar mass are smaller and
therefore the slepton-squark spectrum is typically lighter.

Note that in these scenarios the existence of a very light neutralino
could induce the invisible decay of the lightest Higgs,
$h\to\neut\neut$, thus making Higgs detection more compelling. 
This was studied in Ref.~\cite{invisible}, where some
implications for dark matter were also investigated. 
The branching ratio of the former decay is larger for small values of
the $\mu$ parameter. In this respect it is interesting to point out 
that in 
case
a) $\mu$ can be very efficiently decreased
and thus lead to a large reduction of the visible
Higgs decay rates.

Departures from the case $\delta'_2=\delta'_3$ will 
affect the size of the allowed regions in the parameter space due
to the effect of the experimental constraints. Once more, if
$\delta'_2\ll\delta'_3$ the experimental bound on the chargino might
not be satisfied for small values of $M$, thus excluding those
regions with the lightest neutralinos. Also if
$\delta'_2\gg\delta'_3$ the lower limit on $\asusy$ and the \bsg\
constraint may exclude the whole parameter space.

Let us finally remark that if non-universalities in the Higgs sector
were chosen with the opposite sign for the $\delta$ parameters 
with respect to those in \eq{nunivsc}, i.e., $\delta_1>0$ and
$\delta_2<0$, 
then the value of $m_A$ would increase with respect to its value with
universal scalars. As a consequence, no reduction in the relic density
of these light neutralinos would be obtained and $\relic$ would
exceed its upper limit.

\section{Conclusions}
\label{conclusions}

In this paper we have analysed the theoretical predictions for
neutralino dark matter direct detection in the context of a 
SUGRA theory where both the scalar and
gaugino soft supersymmetry-breaking terms have a non-universal
structure. More specifically, we have computed the predictions for the
scalar neutralino-nucleon
cross section and compared it with the sensitivity of dark matter
detectors. Recent experimental and astrophysical constraints have been
taken into account in the calculation, as well as those derived from
the absence of charge and colour breaking minima.

Gaugino non-universalities are complementary to those in the scalar
masses, allowing more flexibility in
the neutralino sector. 
This is due to the freedom to play with
the value of $M_1$, which is not subject to such strict constraints as
$M_2$ (which is constrained by $\asusy$ and the experimental 
bound on the chargino mass) 
and $M_3$ (whose value is limited by
the lower bound on the Higgs mass and the value of \bsg).
In particular,
neutralinos in the detection range can be
obtained with a wide range of masses.
We have illustrated this possibility by applying gaugino
non-universalities on examples with non-universal scalars which lead
to large predictions for the neutralino-nucleon cross section.

On the one hand, if 
the value of 
$M_1$ is increased with respect to
$M_{3}$ heavier neutralinos are found with a slight increase in their
detection cross section, due to the enhancement of their Higgsino
components.
In this sense, 
neutralinos with a mass as heavy as about $400$ GeV can be obtained 
with a large cross section
($\crosssec\gsim10^{-6}$ pb),
for moderate and
large values of $\tan\beta$.
The increase in $M_1$ is limited by the fact
that a purely wino or Higgsino leads to a very
important decrease in the relic density and is therefore inconsistent
with the astrophysical bounds.

On the other hand, decreasing $M_1$ with respect to $M_3$
light 
neutralinos, with a more important bino composition, can be
obtained. Although, 
due to the increase in the $\mu$ parameter, their cross section is
typically smaller, compatibility with the DAMA region can still be
obtained. For instance, $\crosssec\gsim10^{-6}$ pb is possible with
$\neumass\sim100$ GeV.

Finally, very light neutralinos ($\neumass\lsim M_Z/2$)
can appear for $M_1\ll M_{2,3}$
with a detection cross
section near the sensitivity of present dark matter detectors and
compatible with the DAMA region.
In order to obtain such light neutralinos the presence of
non-universal scalars which lead to a very light CP-odd Higgs is
crucial, combined with a moderate or large hierarchy in the gaugino
sector. 
For example, neutralinos as light as $30$ GeV and $15$ GeV 
can be obtained with
$M_1=4\, M_{2,3}$ and $M_1=11\, M_{2,3}$, respectively.
This is therefore a possibility that is neither present with just
non-universal scalars, where the lower bound on the neutralino mass
is due to the lower bound on the common gaugino mass, $M$, nor
with just non-universal gauginos, where the reduction in $m_A$ cannot
be achieved.

This general analysis can be very useful in the study of more
specific cases, such as the supergravity theories resulting at the
low energy limit
of string constructions. In particular, D-brane scenarios in
Type I string theory give rise to theories where non-universalities
appear both in the scalar and gaugino sectors.

\vspace*{1cm}
\noindent{\bf Acknowledgements}

We would like to thank 
M.~E.~G\'omez 
for useful discussions and comments.
The work of D.G. Cerde\~no was supported in part by the Deutsche
Forschungsgemeinschaft, 
the DAAD, 
and the European Union under contract
HPRN-CT-2000-00148.
The work of C. Mu\~noz was supported 
in part by the Spanish DGI of the
MCYT under Acci\'on Integrada Hispano-Alemana HA2002-0117,
and 
under contracts BFM2003-01266 and FPA2003-04597;
and the European Union under contract 
HPRN-CT-2000-00148.

\providecommand{\href}[2]{#2}

\end{document}